\documentclass[11pt,a4paper]{article}
\usepackage{jheppub}
\usepackage{amsmath,amssymb,amsfonts,graphicx,tensor}
\usepackage{hyperref}
\hypersetup{
    colorlinks=true,
    linkcolor=blue,
    filecolor=magenta,
    urlcolor=cyan,
    pdftitle={Your Title},
    pdfpagemode=FullScreen,
}

\usepackage{subcaption}
\usepackage{svg}
\usepackage{cleveref}

\usepackage{enumitem}
\usepackage{braket}

\title{ Exploring critical behavior of thermodynamic variables of the Kerr-Newman-AdS black hole in the restricted phase space}
\author{Pabitra Tripathy}
\affiliation{Theory Division, Saha Institute of Nuclear Physics, 
		1/AF, Bidhannagar, Kolkata 700064, India}
\affiliation{
		Homi Bhabha National Institute, Anushaktinagar, 
		Mumbai, Maharashtra 400094, India}

\emailAdd{pabitra.tripathy@saha.ac.in} 

\abstract{The present work investigates the thermodynamic properties of the four-dimensional Kerr-Newmann-AdS black hole, utilizing the recently proposed framework of restricted phase space thermodynamics. This approach introduces a novel set of paired thermodynamic variables: the central charge ($C$) of the corresponding dual conformal field theory and the chemical potential ($\mu$). Through rigorous analysis, we establish fundamental relationships, including the Euler relation, Gibbs-Duhem relation, and the zeroth-order homogeneity of intensive variables. Furthermore, numerical techniques are employed to explore thermodynamic processes between the conjugate variables in canonical and grand canonical ensembles. Our investigation reveals first-order and second-order phase transitions across various macroscopic processes. Despite the absence of complete analytical expressions, our findings unveil striking similarities in behavior between Reissner-Nordstrom-Anti-de Sitter and Kerr-Anti-de Sitter, underscoring the presence of underlying universality within the restricted phase space thermodynamics formalism.}
\begin{document}

\maketitle
\newpage
\newpage

\section{Introduction}
The idea that black holes act like thermodynamic systems emerged from studies in black hole mechanics, where similarities to regular thermodynamics became apparent \cite{Bardeen:1973gs}. Bekenstein took this further by suggesting that a black hole's entropy is linked to its surface area \cite{PhysRevD.7.2333}. Initially, classical black holes were thought not to reach thermal equilibrium because they don't emit radiation. But Hawking's groundbreaking work changed that, opening up a new field of black hole thermodynamics \cite{Hawking:1975vcx}. Understanding black hole thermodynamics is crucial for bridging the gap between two fundamental theories: general relativity and quantum mechanics. AdS black holes have properties different from those in asymptotically flat space. For example, they have a minimum temperature, can undergo phase transitions, and exhibit canonical ensembles. Early studies by Hawking and Page revealed a significant phase transition in AdS black holes at a critical temperature \cite{Hawking:1982dh}. Subsequent research, including work by Witten within the AdS/CFT framework, has further explored these phenomena across different types of black holes \cite{witten1998antide, Birmingham_1999, Emparan_1999, Chamblin_1999, Louko_1996, Mitra_1998, Pe_a_1999}. Caldarelli and colleagues were among the first to delve into the thermodynamic behavior of Kerr-Newman-AdS black holes, uncovering intriguing phase transitions tied to factors like mass, charge, and rotation \cite{Caldarelli_1999}.

A recent advancement in black hole thermodynamics is known as extended phase space thermodynamics, often referred to as black hole chemistry \cite{Kastor_2009, Dolan_2011, Dolan1_2011, Cai_2013, Kubiz_k_2017}. This approach originated from recognizing the negative cosmological constant $\Lambda$ as a form of thermodynamic pressure. This addition introduces a volume term for black holes, which is conjugate to the thermodynamic pressure. The novelty of the EPST formalism lies in incorporating a pressure-volume term into the thermodynamic phase space, mirroring the structure of conventional thermodynamics \cite{Kastor_2009}. Consequently, several intriguing features emerged, such as AdS black holes exhibiting phase transitions akin to those seen in Van der Waals fluids \cite{Chamblin_1999, KUBZANAK}, the phenomenon of reentrant phase transitions \cite{Altamirano_2013}, and the utilization of black holes as heat engines \cite{Johnson_2014}, along with the existence of triple points \cite{Altamirano_2014, Wei_2014}. Recently, using holographic braneworlds, researchers demonstrated how the dynamic cosmological constant naturally arises on the brane \cite{Frassino_2023}. Visser further expanded the extended phase space thermodynamics by introducing a new set of conjugate thermodynamic variables: central charge $C$ and chemical potential $\mu$ \cite{Visser_2022}. While central charge and chemical potential have been previously considered in other works \cite{Kastor_2014, Karch_2015, Maity_2017}, Visser's contribution lies in addressing the thermodynamic properties of both the bulk and the CFT on the boundary. Subsequently, Ahmed et al. established the holographic duality of the first law of extended thermodynamics \cite{ahmed2023holographic}. This law reflects how changes in $\Lambda$ in the bulk correspond to alterations in central charge and volume in the CFT on the boundary, ultimately aligning with the principles of restricted phase space thermodynamics \cite{Gao_2022}. Unlike the restricted phase space formalism, where the cosmological constant is held constant while Newton's constant remains fixed, in extended thermodynamics, the authors treated the cosmological constant as a variable.

Despite the various successes of extended thermodynamics, it still grapples with a couple of challenges, namely homogeneity and the ``ensemble of theories" issue. In standard thermodynamics, internal energy behaves as a first-order homogeneous function of extensive variables. However, in EPST, we observe that internal energy functions as a homogeneous function of extensive variables but with a different order in the Euler relation. Another concern relates to the variation of the cosmological constant, which is inherently tied to a specific theory of gravity. Altering $\Lambda$ entails navigating through a spectrum of gravity models, thereby highlighting what is known as the ensemble of theories problem. While Visser's work manages to sidestep the homogeneity issue, it still contends with the ensemble of theories challenge.

While preserving many valuable outcomes of EPST, a recent advancement known as restricted phase space thermodynamics has emerged to address existing discrepancies \cite{Gao_2022}. This novel approach involves a subtle modification of extended-phase space thermodynamics. Unlike its predecessor, restricted phase space thermodynamics does not treat the cosmological constant as a variable. Instead, it expands the phase space of conventional black hole thermodynamics by introducing Newton's constant as a variable. In addition to the standard conjugate variables of black hole thermodynamics, this new framework introduces a pair of conjugate thermodynamic variables: the central charge $C$ and the chemical potential $\mu$. According to holographic duality, the central charge corresponds to the effective number of microscopic degrees of freedom within the bulk. Notably, the restricted phase space formalism resolves the ensemble of theories issue.

 It is necessary to explain the different roles of Newton's constant and the cosmological constant. In spacetime with no matter field, the Einstein-Hilbert action in its usual form,
 \begin{equation}
     S=\frac{1}{16\pi G}\int(R-2\Lambda)\sqrt{-g}d^4x
 \end{equation}
 shows that $\Lambda$ is a part of Lagrangian density. Varying $\Lambda$  leads to a change in the equation of motion, resulting in a different set of gravitational field equations and a different geometry that solves these equations. Thus, varying the cosmological constant alters the underlying gravity theory. This means that varying $\Lambda$ represents black holes from different gravity models rather than the same model.\\
 In contrast, Newton's constant $G$ is an overall factor in front of the total action. Varying 
$G$ effectively rescales the action without changing the field equations or the geometry of the spacetime. For instance, in the nonvacuum case, adding a matter field to the Lagrangian only changes the proportionality factor between geometry and matter. Rescaling $G$ requires rescaling integration constants, such as $Q$ in Einstein-Maxwell theory. Therefore, varying $G$ does not alter the underlying gravity theory.\\
 Varying AdS curvature corresponds to the black hole states with the variable cosmological constant, which implies an alternation of gravity theories. Such processes are meaningful on the CFT side but not on the gravity side.\\
It is also important to clarify that although variables $\mu$ and $C$ are borrowed from the dual CFT, they can be interpreted as the chemical potential and the effective number of microscopic degrees of freedom ($N_{bulk}$) of the black hole in the bulk. We study the thermodynamics of the black hole in the bulk, not in the dual CFT. For simplicity, we use ($\mu,C$),  but they can be replaced with ($\mu_{bulk}, N_{bulk}$). To put it another way, in the holographic dictionary, one can add a new rule, $\mu_{CFT}=\mu_{bulk},\, C=N_{bulk}$.

Previous studies have extensively examined the first-order phase transition and the Hawking-Page transition in restricted phase space formalism for RN-AdS black holes in Einstein-Maxwell theory \cite{Gao_2022}. Our work follows a similar trajectory, focusing on the study of restricted phase space formalism for KN-AdS black holes. While four-dimensional Kerr-Newman-AdS black holes have been previously investigated in extended phase space and standard phase space, our study aims to explore their critical behaviors using restricted phase space formalism. By employing numerical analysis, we derive critical parameters and identify a more suitable fitting function. Comparisons with RN-AdS and Kerr black holes under appropriate limits reveal similar thermodynamic phase structures, mirroring those of non-charged and non-rotating cases. The insights garnered from this analysis contribute to a deeper understanding of restricted phase space formalism.

Our current analysis shares certain aspects with recent investigations that are based on extended-phase space thermodynamics \cite{Gong_2023, Cong_2022}. Although our approach is technically similar in many ways, the physical interpretation of EPS significantly differs from ours. In EPS, the cosmological constant varies in the bulk while the gravitational constant remains fixed, leading to the inclusion of a pressure-volume term in the first law of thermodynamics. The negative cosmological constant is interpreted as bulk pressure, $P=-\frac{\Lambda}{8\pi G}$. This interpretation has some peculiar features, one of which is identifying the black hole's mass with enthalpy rather than the system's energy. Considering the central charge and its conjugate chemical potential later addressed several key issues with the EPS formalism. In our study, we examine the thermodynamics of a rotating charged black hole using the restricted phase space formalism. The main difference from the EPS formalism is that we do not vary the cosmological constant to avoid altering the theory. Instead, we treat the gravitational constant as a variable.

The paper is structured as follows:
\begin{itemize}
    \item In Section 3, we provide a brief overview of the restricted phase space formalism, addressing the issues encountered in EPST and proposing solutions to mitigate these challenges.
\end{itemize}
\begin{itemize}
    \item Section 4 offers a concise review of the RPS formalism applied to the RN-AdS black hole.
\end{itemize}
\begin{itemize}
    \item Continuing the discussion, Section 5 extends the review to include the application of the RPS formalism to the Kerr-AdS black hole.
\end{itemize}
\begin{itemize}
    \item Section 6 delves into the examination of the RPS formalism for the charged rotating black hole within Einstein's theory. This section includes discussions on the equation of state, homogeneity behavior, and critical parameters, as well as the presentation of various plots such as $T-S$, $F-T$, $\Omega-J$, and $\mu-C$. These plots aid in the analysis of different thermodynamic processes.
\end{itemize}
\begin{itemize}
    \item Finally, in Section 7, we conclude the paper by summarizing the key findings and discussing potential future research directions.
\end{itemize}

\section{Review of RPST formalism}
As its name implies, the RPST formalism arose from a restriction of the well-established EPST or black hole chemistry. Therefore, it is essential to provide a brief overview of the EPST formalism to maintain the self-contained nature of this article. In this section, we outline the key features of the EPST formalism, highlight its associated issues, and demonstrate how the RPST formalism partially resolves these challenges. Our discussion closely aligns with the insights presented in the work by Gao et al. \cite{Gao_2022}.

In the framework of EPST, the expansion of the traditional black hole phase space arises from the incorporation of a pair of conjugate variables: pressure and volume. The pressure term is directly linked to the negative cosmological constant through the equation $P=-\frac{\Lambda}{8\pi G_D}$, where $\Lambda$ represents the negative cosmological constant, and $G_D$ denotes the $D$-dimensional gravitational constant. For asymptotically Anti-de Sitter black holes, the cosmological constant can be expressed as $\Lambda=-\frac{(D-1)(D-2)}{2l^2}$, where $l$ denotes the AdS radius.

The first law of black hole thermodynamics, when considering a negative cosmological constant, takes on a generalized form as follows
\begin{equation}\label{eq1}
    \delta M=T \delta S+V \delta P+\Omega \delta J+ \Phi \delta Q.
\end{equation}
Here, $T$, $S$, $\Omega$, $J$, $\Phi$, and $Q$ represent the black hole's temperature, entropy, angular velocity, angular momentum, electric potential, and electric charge, respectively. The term $V$ denotes the thermodynamic volume conjugated to the thermodynamic pressure $P$, defined by $V=\left(\frac{\partial M}{\partial P}\right)_{S,Q,J}$. In comparison to the first law of ordinary thermodynamics, the quantity $M$ no longer solely represents the black hole's mass but is termed enthalpy. It encompasses the total energy of the black hole, including both its internal energy and the energy required to displace the vacuum energy.

Additionally, the corresponding Smarr relation for a $D$-dimensional charged, rotating black hole within the EPST framework is expressed as
\begin{equation}\label{eq2}
    M=\frac{D-2}{D-3}(TS+\Omega J)+\Phi Q-\frac{2}{D-3}PV.
\end{equation}

This framework (EPST) faces several issues, which are as follows:
\begin{itemize}
    \item The first issue is that EPST allows the pressure $P$ to vary. Varying $P$ is associated with variations in $\Lambda$, which serves as a defining parameter for the model. Consequently, changes in $P$ correspond to shifts between different gravity models. This aspect of the framework implies the existence of an ensemble of gravity theories.
\end{itemize}
\begin{itemize}
    \item $M$ is not the total energy of the black hole; rather, it represents the enthalpy. In traditional black hole thermodynamics, the black hole mass is interpreted as the total energy of the black hole spacetime. Therefore, EPST's convention contradicts this traditional interpretation.
\end{itemize}
\begin{itemize}
    \item The meaning of the thermodynamic volume $V$ remains unclear.
\end{itemize}
\begin{itemize}
    \item From equation (\ref{eq1}), one derives the following relationships:
    \begin{equation}
        T=\left(\frac{\partial M}{\partial S}\right)_{J,Q,P},\,\,\,\,\Omega=\left(\frac{\partial M}{\partial J}\right)_{S,Q,P},\,\,\,\,\Phi=\left(\frac{\partial M}{\partial Q}\right)_{S,J,P}, \,\,\,\, V=\left(\frac{\partial M}{\partial P}\right)_{S,Q,J}.
    \end{equation}
Using these identities, the Smarr relation (\ref{eq2}) can be expressed as:
 \begin{equation}
        M=\frac{D-2}{D-3}\left[S\left(\frac{\partial M}{\partial S}\right)_{J,Q,P}+J\left(\frac{\partial M}{\partial J}\right)_{S,Q,P}\right]+Q\left(\frac{\partial M}{\partial Q}\right)_{S,J,P}-\frac{2}{D-3}V\left(\frac{\partial M}{\partial P}\right)_{S,Q,J}.
    \end{equation}
This equation suggests that $M$ behaves as a homogeneous function of $S$ and $J$ with an order of $\frac{D-2}{D-3}$, of $Q$ with an order of 1, and of $V$ with an order of $\frac{2}{D-3}$. In contrast, in standard thermodynamics, thermodynamic potentials must be homogeneous functions of the related additive extensive variables of the first order. If we examine Euler's relation, the coefficients appearing on the right-hand side are all equal to one. This disparity represents a significant inconsistency between traditional black hole thermodynamics and extended phase space thermodynamics.

\end{itemize}
In a recent work by Visser \cite{Visser_2022}, the issue is elegantly addressed and resolved by introducing the central charge as a new thermodynamic variable. The central charge $C$ corresponds to the central charge of the holographic dual theory, defined as $C=\frac{l^{D-2}}{G_D}$. With this introduction, we can now redefine the function $M(A, J, Q,\Lambda, G_D)$. Then we can write,
\begin{equation}\label{eq3}
\begin{split}
dM =  \frac{\kappa}{2\pi}d\left(\frac{A_{D-2}}{4G_D}\right) + \Omega dJ + &\frac{\Phi}{l}d(Ql) 
 - \frac{M}{D-2}\frac{dL^{D-2}}{L^{D-2}} + \\
 &\left(M-\frac{\kappa A_{D-2}}{8\pi G_D}-\frac{\Phi}{l}Ql-\Omega J\right)\frac{d(l^{D-2}/G_D)}{l^{D-2}/G_D}.
\end{split}
\end{equation}
Here, $A$ represents the area of the event horizon, and $\kappa$ stands for the surface gravity.

Introducing the definitions, $E=M,\,\,T=\frac{\kappa}{2\pi},\,\,S=\frac{A_{D-2}}{4G_D}, \,\,\Bar{\Phi}=\frac{\Phi}{l},\,\,\Bar{Q}=Ql\,\, \text{and} \,\,V=L^{D-2}$ and utilizing the equation of state , $E=(D-2)PV$, equation (\ref{eq3}) can be reformulated as
\begin{equation}
    dE=TdS+\Omega dJ+\Bar{\Phi}d\Bar{Q}-PdV+\mu dC.
\end{equation}
Here, $\mu$ is determined through the identity given by
\begin{equation}\label{eq4}
    E=TS+\Bar{\Phi}\Bar{Q}+\Omega J+\mu C.
\end{equation}
This equation bears resemblance to the Euler relation found in standard thermodynamics, with the exception of the $PV$ term. Here, $V$ is redefined as the spatial size of the dual CFT. The $C$-charge serves as a gauge of the microscopic degrees of freedom within the CFT, while $\mu$, its conjugate, signifies the chemical potential. However, it's important to note that $V$ is defined in terms of $\Lambda$, thus variations in $V$ correspond to variations in $\Lambda$. Consequently, the formulation introduced by Visser does not entirely alleviate the ensemble of theories issue.

To address this issue, Gao and Zhao \cite{Gao_2022} proposed the RPST using a straightforward approach. By constraining the variation of the cosmological constant, they effectively stabilized the gravity theory. Consequently, the $PV$ term vanishes from the first law. Thus, within the RPST framework, Equation (\ref{eq1}) can be adjusted as follows:
\begin{equation}\label{eq5}
     \delta M=T \delta S+\Omega \delta J+ \Bar{\Phi} \delta \Bar{Q}+\mu dC.
\end{equation}
Additionally, the Smarr formula can be expressed as follows:
\begin{equation}
    M=TS+\Bar{\Phi}\Bar{Q}+\Omega J+\mu C.
\end{equation}
In this formalism, while the variation of the cosmological constant is restricted, it permits the variation of the gravitational constant $G$.

\section{Review of thermodynamics for  Reissner-Nordstrom-AdS black hole in the restricted phase space }
Here, we provide a brief overview of the application of the RPST formalism to the Reissner-Nordstrom-AdS black hole and examine its critical behavior. We adhere to the methodology outlined in \cite{Gao_2022}. This concise review serves a dual purpose: firstly, it lays the groundwork for comparing our primary findings concerning the Kerr-Newmann-AdS black hole with those presented here under certain limits. Secondly, it aids readers in grasping the background information, ensuring a smooth flow throughout the paper. It's important to note that throughout this paper, we confine our analysis to the four-dimensional case.

 The metric for the spherical Reissner-Nordstrom-AdS black hole and its corresponding electromagnetic potential can be expressed as follows
 \begin{equation}
 \begin{split}
      ds^2=&-f(r)dt^2+\frac{dr^2}{f(r)}+r^2(d\theta^2+\sin^2\theta d\phi^2),\\
      &A^\mu=\left(-\frac{Q}{r}\right).
\end{split}
\end{equation}
Where,
\begin{equation}\label{eq6}
    f(r)=1-\frac{2GM}{r}+\frac{GQ^2}{r^2}+\frac{r^2}{l^2}.
\end{equation}
The equation $f(r)=0$ yields two real roots, denoted as $r_\pm$, where $r_+$ represents the radius of the event horizon. Expressing the black hole mass using equation (\ref{eq6}), we obtain
\begin{equation}\label{eq7}
    M=\frac{r_+}{2G}\left(1+\frac{r^2_+}{l^2}+\frac{GQ^2}{r_+^2}\right).
\end{equation}
The set of conjugate variables can be described as follows:
\begin{itemize}
    \item The entropy and temperature of the black hole are represented by
\begin{equation}\label{eq8}
    S=\frac{A}{4G}=\frac{\pi r_+^2}{G},\,\,\,\,\, T=\frac{f'(r_+)}{4\pi}=\frac{1}{2\pi r_+}\left(\frac{GM}{r_+}-\frac{GQ^2}{r^2_+}+\frac{r^2_+}{l^2}\right).
\end{equation}
\end{itemize}
\begin{itemize}
    \item The rescaled electric charge ($\Bar{Q}$) and its conjugate, the electric potential ($\Bar{\Phi}$), can be expressed as
    \begin{equation}\label{eq9}
      \Bar{Q}=\frac{Ql}{\sqrt{G}}, \,\,\,\,\,\,\, \Bar{\Phi}=\frac{Q\sqrt{G}}{r_+l}.
    \end{equation}
\end{itemize}
\begin{itemize}
    \item The central charge ($C$) and its conjugate variable ($\mu$), representing the chemical potential, are given by
    \begin{equation}\label{eq10}
        C=\frac{l^2}{G},\,\,\,\,\,\,\mu=\frac{M-TS-\Bar{\Phi}\Bar{Q}}{C}.
    \end{equation}
\end{itemize}
It is straightforward to demonstrate that
\begin{equation}\label{eq11}
    dM=TdS+\Bar{\Phi}d\Bar{Q}+\mu dC.
\end{equation}
From the expression of $\mu$ given in equation (\ref{eq10}), we can derive the Euler relation as follows
\begin{equation}\label{eq12}
    M=TS+\Bar{\Phi}\Bar{Q}+\mu C.
\end{equation}

The Gibbs-Duhem relation can be derived using equations (\ref{eq11}) and (\ref{eq12}),
\begin{equation}
    d\mu=-\Bar{q}d\Bar{\Phi}-\Bar{s}dT.
\end{equation}
Where $q=\frac{\Bar{Q}}{C}$ and $s=\frac{S}{C}$. It's worth noting that this relation is absent in both traditional and extended-phase space thermodynamics.
 \subsection{Equation of state}
The mass $M$ can be expressed as a function of state variables $S,C\,\,\, \text{and}\,\,\,\Bar{Q}$ in the following manner
\begin{equation}
    M(S,\Bar{Q},C)=\frac{S^2+\pi S C+\pi^2\Bar{Q}^2}{2\pi^{3/2}l\sqrt{SC}}.
\end{equation}
The equations of states are provided as follows
\begin{equation}\label{eqT}
    \begin{split}
         T=\left(\frac{\partial M}{\partial S}\right)_{\Bar{Q},C}=\frac{3S^2+\pi S C-\pi^2\Bar{Q}^2}{4\pi^{3/2}lS\sqrt{SC}},
    \end{split}
\end{equation}
\begin{equation}\label{phibar}
    \Bar{\Phi}=\left(\frac{\partial M}{\partial Q}\right)_{S,C}=\sqrt{\frac{\pi}{SC}}\frac{\Bar{Q}}{l},
\end{equation}
\begin{equation}\label{muequation}
    \mu=\left(\frac{\partial M}{\partial C}\right)_{S,\Bar{Q}}=-\frac{S^2-\pi S C+\pi^2\Bar{Q}^2}{4\pi^{3/2}lC\sqrt{SC}}.
\end{equation}
\subsection{Thermodynamic process}
In the equations of state, the interrelationship between three intensive variables $(T,\Bar{\Phi},\mu)$ and three extensive variables $(S,\Bar{Q}, C)$ is elucidated. We specifically investigate thermodynamic processes involving one intensive and one extensive variable, while holding the remaining two extensive variables constant.

\subsubsection{\textit{\textit{T-S process at fixed} \textit{$\Bar{Q},C$}}}\label{TS PROCESS RN}
The critical point on the $T-S$ curve at fixed $\Bar{Q},C$  is determined through solution of the following equations
\begin{equation}\label{critical1}
    \left(\frac{\partial T}{\partial S}\right)_{\Bar{Q},C}=0,\,\,\,\,\,\,\,\,\,\,\left(\frac{\partial^2 T}{\partial S^2}\right)_{\Bar{Q},C}=0.
\end{equation}
The critical values of $\Bar{Q}, S$ and $C$ are given by
\begin{equation}\label{criticalrn}
    \Bar{Q_c}=\frac{C}{6},\,\,\,\,\,S_c=\frac{\pi C}{6},\,\,\,\,\, T_c=\frac{\sqrt{6}}{3\pi l}.
\end{equation}
We define the dimensionless relative parameters as follows
\begin{equation}
    q=\frac{\Bar{Q}}{\Bar{Q}_c},\,\,\,\,\,\,s=\frac{S}{S_c},\,\,\,\,\,\,\,\tau=\frac{T}{T_c}.
\end{equation}
Subsequently, the equation of state can be reformulated in terms of these relative parameters, thereby enhancing its conciseness
\begin{equation}
    t=\frac{3s^2+6s-q^2}{8s^{3/2}}.
\end{equation}

\begin{figure}
    \centering
    \begin{subfigure}[b]{0.45\textwidth}
    \centering
    \includegraphics[width=\textwidth]{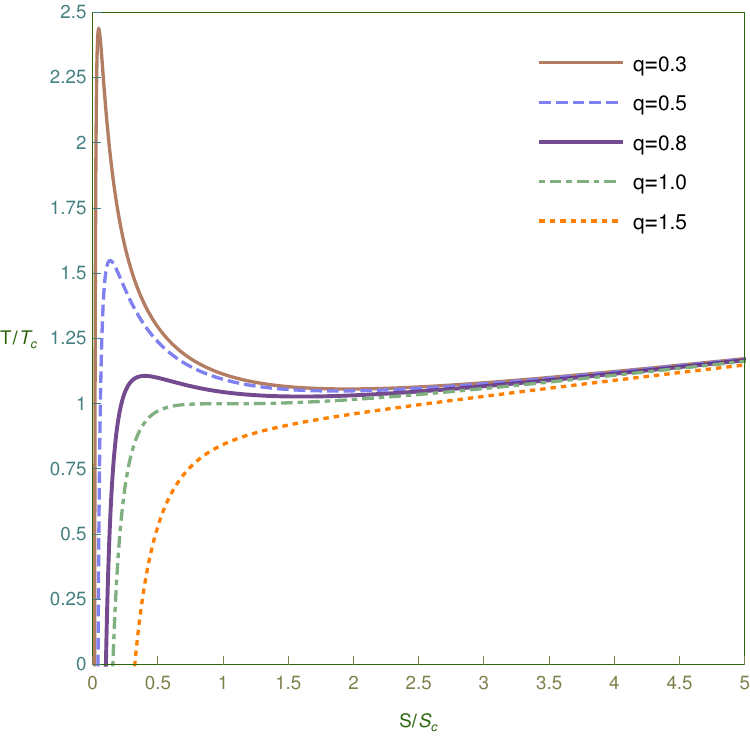}
         \caption{}
         \label{fig: TS RN}
     \end{subfigure}
     \hfill
     \begin{subfigure}[b]{0.48\textwidth}
    \centering
    \includegraphics[width=\textwidth]{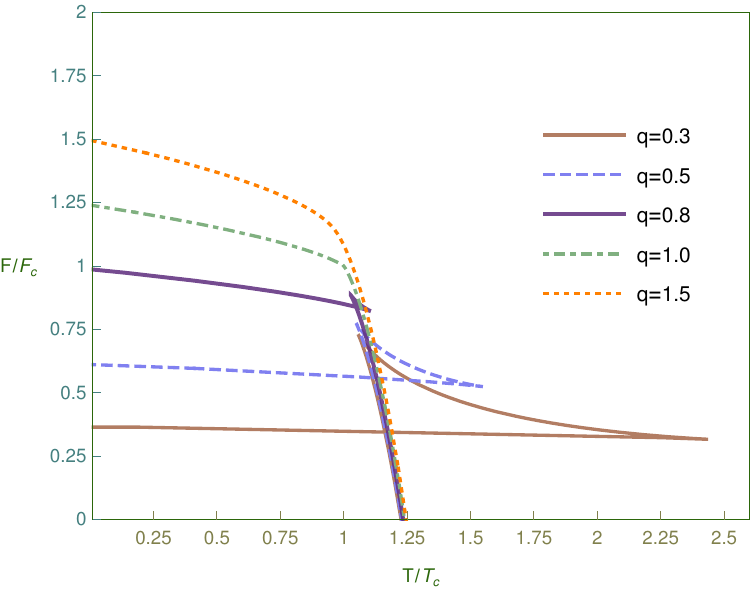}
    \caption{}
     \label{fig: FT RN}
     \end{subfigure}

     \caption{The T-S and F-T curves at constant $\Bar{Q}$.}
     \label{fig: TSFT RN}
\end{figure}

\subsubsection{\textit{F-T process at fixed} $\Bar{Q},C $ }
The Helmholtz free energy can be expressed as
\begin{equation}\label{critf}
    F(T,\Bar{Q},C)=M(T,\Bar{Q},C)-TS.
\end{equation}
The critical value of the Helmholtz free energy can be determined by substituting all the critical values from equation (\ref{criticalrn}) into equation (\ref{critf}), yielding
\begin{equation}\label{fcrnlimit}
    F_c=\frac{\sqrt{6}C}{18l}.
\end{equation}
 Introducing the relative parameter $f=\frac{F}{F_c}$,  equation (\ref{critf}) can be elegantly reformulated as
\begin{equation}
    f=\frac{q^2+s^2+6s-4\tau s^{3/2}}{4s^{1/2}}.
\end{equation}
The expressions of $t$ and $s$ distinctly reveal that owing to their scaling properties, both $t$ and $s$ are independent of the central charge $C$.

For the various values of $\Bar{Q}$ the behavior of $T-S$ and $F-T$ curves are depicted in \ref{fig: TS RN} and \ref{fig: FT RN} respectively.  In figure \ref{fig: TSFT RN}, it is evident that for $\Bar{Q}<\Bar{Q}_c$ the $T-S$ curve exhibits non-monotonicity and the $F-T$ curves display swallowtail behavior. Conversely when, $\Bar{Q}>\Bar{Q}_c$, both the non-monotonicity and swallowtail behavior vanish. This observation indicates a first-order van der Waals-like phase transition below $\Bar{Q_c}$ with no phase transition occurring above the subcritical value of $Q$. The intersection point of the swallowtail structure emerges at some critical temperature $T>T_c$.

\begin{figure}
    \centering
     \centering
    \includegraphics[width=0.60\textwidth]{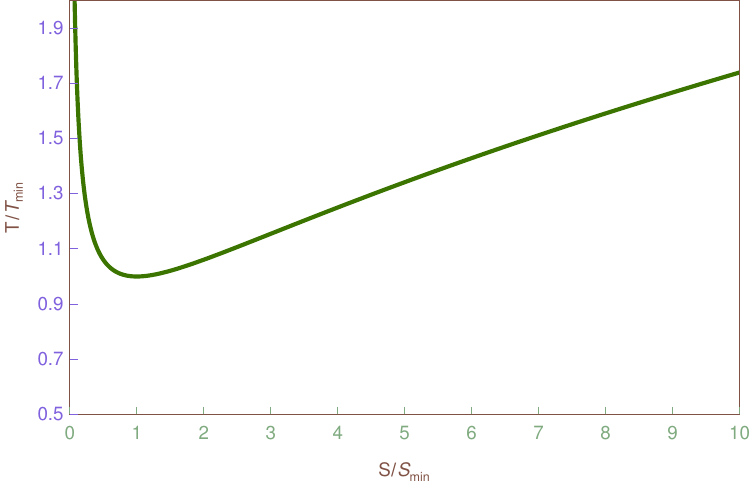}
      \caption{T-S curve at constant $\Bar{\Phi}$.}
     \label{fig: TSFT RN}
\end{figure}

\subsubsection{\textit{T-S process at fixed} $\Bar{\Phi},C$}\label{Fixed phi TS process}
Utilising equations (\ref{eqT}) and (\ref{phibar}), $T(S,C,\Bar{\Phi})$ can be expressed as
\begin{equation}\label{tsphi}
    T(S,C,\Bar{\Phi})=\frac{3S+\pi C(1-l^2\Bar{\Phi}^2)}{4\pi^{3/2}l\sqrt{SC}},
\end{equation}
The condition $T\geq 0$, imposes a constraint on $\Bar{\Phi}$ namely,
\begin{equation}
    \Bar{\Phi}\leq \frac{1}{l}
\end{equation}
For a fixed $C$, and $\Bar{\Phi}$, equation (\ref{tsphi}) reveals that $T$ attains a minimum value
\begin{equation}\label{SMINTMINRN}
    S_{min}=\frac{\pi C(1-l^2\Bar{\Phi}^2)}{3},\,\,\,\,\,\,\, T_{min}=\frac{\sqrt{3(1-l^2\Bar{\Phi}^2)}}{2\pi l}.
\end{equation}
Let's define the relative parameters as follows
\begin{equation}
    \Bar{t}=\frac{T}{T_{min}},\,\,\,\,\,\,\,\,\, \Bar{s}=\frac{S}{S_{min}}.
\end{equation}
In terms of relative parameters, equation (\ref{tsphi}) can be rewritten as
\begin{equation}\label{tsequation}
    \Bar{t}=\frac{\Bar{s}+1}{2\sqrt{\Bar{s}}}.
\end{equation}
Once again we observe that due to scaling, the equation (\ref{tsequation}) remains independent of $C, \Bar{\Phi}$. The $T-S$ iso-voltage curves depicted in Fig. \ref{fig: TSFT RN} illustrate that $T$ can not fall below $T_{min}$. Two black hole states correspond to the same $T,\Bar{\Phi}, C$ for different entropies. Among these states, the one with maximum entropy is deemed the stable state as per the theory of maximum entropy. Following a slight perturbation, the state with lower entropy transitions to the state of higher entropy while maintaining the same $T,\Bar{\Phi}, C$. Although this transition lacks an associated transition temperature, it can be considered a phase transition.

\subsubsection{$\mu-C$ \textit{process at fixed} ($S,\Bar{Q}$) and ($S,\Bar{\Phi}$)}
\begin{figure}
    \centering
     \centering
    \includegraphics[width=0.60\textwidth]{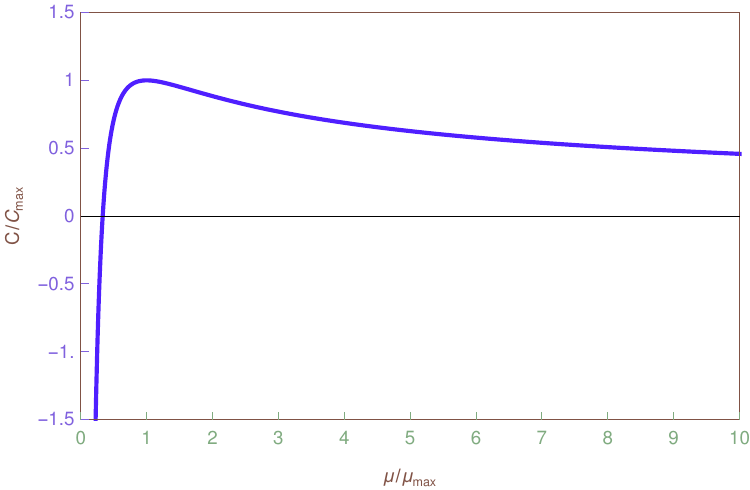}
      \caption{$\mu-C$ curve at constant $\Bar{Q}$ or $\Bar{\Phi}$.}
     \label{fig: muc curve}
\end{figure}

To analyze the $\mu-C$ process, we examine equation (\ref{muequation}) and observe that for a fixed $S,\Bar{Q}$, $\mu$ reaches a maximum at $C=C_{max}$. These are given by
\begin{equation}
    \mu_{max}=\frac{\sqrt{3}S}{18l\sqrt{S^2+\pi^2\Bar{Q}^2}},\,\,\,\,\,\,\,\,\,\,C_{max}=\frac{3(S^2+\pi^2\Bar{Q}^2)}{\pi S}.
\end{equation}
Let's define the dimensionless relative parameter as
\begin{equation}\label{dimns}
    \Bar{\mu}=\frac{\mu}{\mu_{max}},\,\,\,\,\,\,\,\,\,\,\,c=\frac{C}{C_{max}}.
\end{equation}\label{mufin}
In terms of these relative parameters equation (\ref{muequation}) can be rewritten as
\begin{equation}\label{mucrn}
    \Bar{\mu}=\frac{3c-1}{2c^{3/2}}.
\end{equation}
We plot the $\mu-C$ curves in Fig. \ref{fig: muc curve} and observe that they exhibit a similar behavior to the isovoltage $T-S$ curves. It is worth noting that at $C=\frac{1}{3}C_{max}$, both the chemical potential and Gibbs free energy vanish.

Using equations (\ref{muequation}) and (\ref{phibar}) $\mu$ can be expressed in terms of $(S,C,\Bar{\Phi})$ as follows
\begin{equation}
    \mu(S,C,\Bar{\Phi})=-\frac{S^2-\pi S C+\pi S C l^2\Bar{\Phi}^2}{4\pi^{3/2}lC\sqrt{SC}}.
\end{equation}
At constant $C, \Bar{\Phi}$, there exists a maximum value $\mu_{max}$ corresponding to $C_{max}$,
\begin{equation}
    \mu_{max}=\frac{\sqrt{3(1-l^2\Bar{\Phi^2})^3}}{18l},\,\,\,\,\,\,\,\,\,\,\,\,\,\,C_{max}=\frac{3S}{\pi(1-l^2\Bar{\Phi^2})}.
\end{equation}

Utilizing the relative parameters defined in equation (\ref{dimns}) we arrive at equation (\ref{mufin}). Thus, $\mu-C$ curves at fixed $(S,\Bar{Q})$ are identical to those at fixed $(S,\Bar{\Phi})$.

\section{Review of thermodynamics for  Kerr-AdS black hole in the restricted phase space }
Here, we delve into the RPST formalism for the Kerr-AdS black hole, following the methodology outlined in Ref. \cite{Gaok_2022}.

The metric of a four-dimensional Kerr-AdS black hole in Boyer-Lindquist-like coordinate $(t,r,\theta, \phi)$ is described by
\begin{equation}
    ds^2= -\frac{\Delta_r}{\rho^2}\left(dt-\frac{asin^2\theta}{\Sigma}d\phi\right)^2+\frac{\rho^2}{\Delta_r}dr^2+\frac{\rho^2}{\Delta_{\theta}}d\theta^2+\frac{sin^2\theta\Delta_\theta}{\rho^2}\left(adt-\frac{r^2+a^2}{\Sigma}d\phi\right)^2.
\end{equation}
where
\begin{equation}
    \begin{split}
        &\rho^2=r^2+a^2cos^2\theta,\,\,\,\,\,\,\Delta_r=(r^2+a^2)\left(1+\frac{r^2}{l^2}\right)-2Gmr,\\
        &\Sigma=1-\frac{a^2}{l^2},\,\,\,\,\,\,\,\,\Delta_\theta=1-\frac{a^2}{l^2}cos^2\theta.
    \end{split}
\end{equation}
The mass and angular momentum are related to $m$, and $a$ as
\begin{equation}
    M=\frac{m}{\Sigma^2},\,\,\,\,\,\,\,J=\frac{am}{\Sigma^2}.
\end{equation}
By solving $\Delta_r(r_+)=0$, where $r_+$ is the radius of the event horizon, we can express the mass as
\begin{equation}\label{M}
    M=\frac{1}{2\Sigma^2r_+G}\left(r_+^2+a^2+\frac{r_+^4}{l^2}+\frac{a^2r_+^2}{l^2}\right).
    \,\,\,\,
\end{equation}
The pair of conjugate thermodynamic variables are given by
\begin{itemize}
    \item Entropy and Temperature are given by
    \begin{equation}\label{S}
    S=\frac{\pi(a^2+r_+^2)}{G\Sigma},\,\,\,\,\,\,\,\,\,\,\,\,T=\frac{r_+}{4\pi(a^2+r_+^2)}\left(1+\frac{a^2}{l^2}+\frac{3r_+^2}{l^2}-\frac{a^2}{r_+^2}\right).
     \end{equation}
\end{itemize}
\begin{itemize}
    \item Angular momentum and angular velocity are given by
    \begin{equation}\label{J}
   J=\frac{a}{2\Sigma^2r_+G}\left(r_+^2+a^2+\frac{r_+^4}{l^2}+\frac{a^2r_+^2}{l^2}\right) ,\,\,\,\,\,\,\,\,\,\,\,\,\Omega=\frac{a\Sigma}{r_+^2+a^2}+\frac{a}{l^2}.
     \end{equation}
\end{itemize}
\begin{itemize}
    \item Chemical potential and central charge are given by
    \begin{equation}\label{mu}
        \mu=\frac{M-TS-\Omega J}{C},\,\,\,\,\,\,\,\,\,C=\frac{l^2}{G}.
    \end{equation}
\end{itemize}
The consistency of the aforementioned thermodynamic variables with the first law of thermodynamics is readily demonstrable,
\begin{equation}
    dM=TdS+\Omega dJ+\mu C.
\end{equation}
Moreover, it's worth noting that these thermodynamic variables also satisfy the Euler relation,
\begin{equation}
    M=TS+\Omega J+\mu C.
\end{equation}
\subsection{Equation of states}
All intensive variables can indeed be expressed as functions of extensive variables \cite{Gaok_2022}. To achieve this, we undertake the task of replacing  $r_+,a$ and $G$ in terms of $J, M,\text{and}\,\,C$. These are given by
\begin{equation}\label{rescale}
    a=\frac{J}{M},\,\,\,\,\,\,\,\,\,\,G=\frac{l^2}{C},\,\,\,\,\,\,\,\,\,\,r_+=\sqrt{\frac{S(l^2M^2-J^2)}{\pi C M^2}-\frac{J^2}{M^2}}.
\end{equation}
Inserting equations (\ref{rescale}) into equation (\ref{M}), we can rewrite the expression for mass ($M$) as follows
\begin{equation}
    M=\frac{\sqrt{(\pi C+S)(S^3+\pi C(4\pi^2J^2+S^2))}}{2\pi^{3/2}l\sqrt{CS}}.
\end{equation}
Then, the equations of states can be expressed as
\begin{equation}\label{TKERR2}
    T=\left(\frac{\partial M}{\partial S}\right)_{J,C}=\frac{4\pi CS^3+3S^4+\pi^2C^2(S^2-4\pi^2J^2)}{4\pi^2lS^{3/2}\sqrt{C(\pi C+S)(S^3+\pi C(4\pi^2J^2+S^2))}},
\end{equation}
\begin{equation}\label{omegaKERR}
    \Omega=\left(\frac{\partial M}{\partial S}\right)_{S,C}=\frac{2\pi^{3/2}J\sqrt{C(\pi C+S)}}{l\sqrt{S(S^3+\pi C(4\pi^2J^2+S^2))}},
\end{equation}
\begin{equation}
    \mu=\left(\frac{\partial M}{\partial C}\right)_{S,J}=\frac{\pi^2 C^2(4\pi^2J^2+S^2)-S^4}{4\pi^{3/2}lC^{3/2}\sqrt{[S(\pi C+S)(S^3+\pi C(4\pi^2 J^2+S^2))]}}.
\end{equation}
\subsection{Thermodynamic process}
In this section, our focus lies in analyzing thermodynamic processes, specifically the $T-S$, $\Omega-J$, and $\mu-C$ processes. To maintain analytical results, we opt for a slow rotation limit. We enforce this restriction by performing a power series expansion of the expressions for $ M, T,\Omega$ and $\mu$ concerning $J$, considering terms up to $J^2$. The modified equations of states are thus given by
\begin{equation}\label{MKERR}
    M(S,J,C)=\frac{C[\pi C(2\pi^2J^2+S^2)+S^3]}{2\pi^{3/2}l\sqrt{S^3C^3}},
\end{equation}
\begin{equation}\label{TEQ}
    T(S,J,C)=\frac{\pi C S^2+3S^3-6\pi^3CJ^2}{4\pi^{3/2}l\sqrt{CS^5}},
\end{equation}
\begin{equation}\label{omegakerr}
    \Omega(S,J,C)=\frac{2\pi^{3/2}J}{l}\sqrt{\frac{C}{S^3}},
\end{equation}
\begin{equation}\label{muKERR}
    \mu(S,J,C)=\frac{\pi C(2\pi^2J^2+S^2)-S^3}{4\pi^{3/2}l\sqrt{C^3S^3}}.
\end{equation}
\subsubsection{\textit{T-S process at fixed J,C}}\label{TS PROCESS KERR}
\begin{figure}
    \centering
    \begin{subfigure}[b]{0.50\textwidth}
    \centering
    \includegraphics[width=\textwidth]{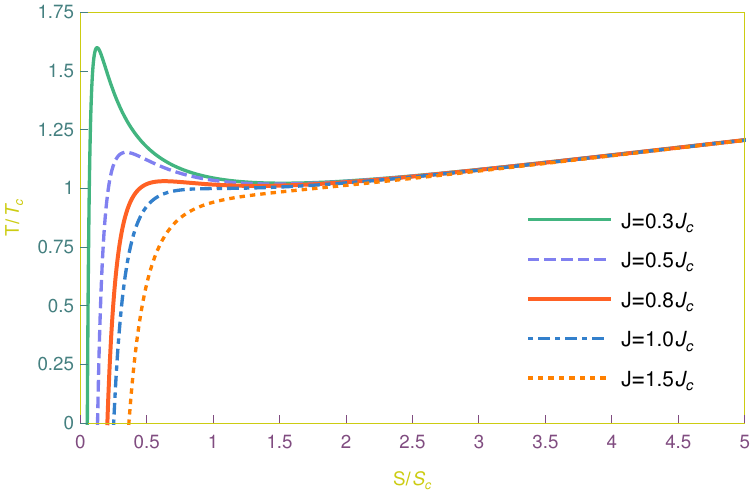}
         \caption{}
         \label{fig: TS KERR}
     \end{subfigure}
     \hfill
     \begin{subfigure}[b]{0.48\textwidth}
    \centering
    \includegraphics[width=\textwidth]{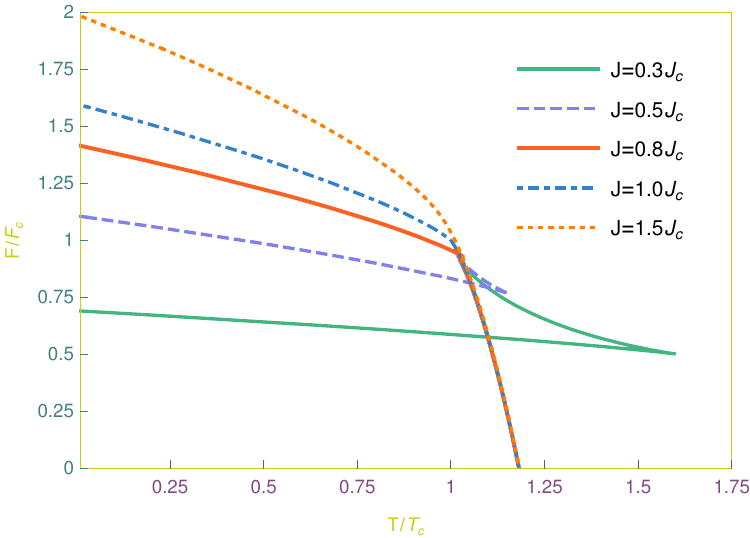}
    \caption{}
     \label{fig: FT KERR}
     \end{subfigure}

     \caption{$T-S$ and $F-T$ curves at fixed $J$}
     \label{fig: TSFT KERR}
\end{figure}

Critical points of the $T-S$ curves are determined by solving the following equations
\begin{equation}\label{critical}
    \left(\frac{\partial T}{\partial S}\right)_{J,C}=0,\,\,\,\,\,\,\,\,\, \left(\frac{\partial^2 T}{\partial S^2}\right)_{J,C}=0.
\end{equation}
Inserting equation (\ref{TEQ}) into equations (\ref{critical}) we obtain the approximated critical parameters as follows
\begin{equation}\label{criticalvalues}
    S_c\approx 0.68250 C,\,\,\,\,J_c\approx 0.02411 C,\,\,\,\,T_c\approx \frac{0.26939}{l}.
\end{equation}
Let's define the relative parameters as
\begin{equation}
    s=\frac{S}{S_c},\,\,\,\,j=\frac{J}{J_c},\,\,\,\,\,t=\frac{T}{T_c}.
\end{equation}
In terms of these relative parameters equation (\ref{TEQ}) can be rewritten as
\begin{equation}\label{tsplot}
    t(s,j)=[(0.41305s+0.63377s)s^2-0.04682j^2]s^{-\frac{5}{2}}.
\end{equation}
We are now equipped to visualize the thermodynamic characteristics of black holes by plotting $T-S$ curves based on the relationship outlined in equation (\ref{tsplot}). In Fig. \ref{fig: TS KERR}, we present these curves, focusing on fixed values of angular momentum ($J$) and central charge ($C$). 

\subsubsection{\textit{F-T process at fixed J, C}}
 The Helmholtz free energy is expressed as
  \begin{equation}\label{helmholtz}
      F(T,C,J)=M(T,C,J)-TS.
  \end{equation}
Utilizing Equations (\ref{MKERR}) and (\ref{TEQ}), we can express the Helmholtz free energy in terms of the variables $C$, $J$, and $S$ as
\begin{equation}\label{FKERR}
    F(S,J,C)=\frac{\pi \sqrt{C^3S^5}(10\pi^2J^2+S^2)-\sqrt{CS^{11}}}{4\pi^{3/2}lCS^4}.
\end{equation}
 By substituting the critical values derived from Equation (\ref{criticalvalues}) into Equation (\ref{FKERR}), we determine the critical value of the free energy, which is expressed as
  \begin{equation}\label{fckerrlimit}
      F_c\approx \frac{0.10556 C}{l}.
  \end{equation}
Introducing the relative parameter $f=\frac{F}{F_c}$, Equation (\ref{helmholtz}) can be reformulated in terms of $j$ and $s$ as follows
\begin{equation}\label{f}
    f(t,j)=[(1.10389-0.23982s)s^2+0.13593j^2]s^{-\frac{3}{2}}.
\end{equation}
We can plot $F-T$ curves using equations (\ref{tsequation}) and (\ref{f}), with $s$ being implicitly replaced. Figure \ref{fig: FT KERR} illustrates these $F-T$ curves, providing a visual representation of the relationship between the Helmholtz free energy ($F$) and temperature ($T$) in Kerr spacetime.

Observing Fig. \ref{fig: TS KERR}, it's evident that the $T-S$ curves exhibit a non-monotonic behavior. Additionally, from Fig. \ref{fig: FT KERR}, the $F-T$ curves display a swallowtail shape below a critical angular momentum range $0<J<J_c$. This characteristic suggests a first-order phase transition in the $T-S$ process at fixed $C$ and $J$. The transition temperature of this phase transition can be determined from the crossing position of the swallowtail. 

\subsubsection{\textit{T-S and F-T process at J=0}}\label{Schwarzschild AdS}
In the limit where $J=0$, we arrive at the Schwarzschild-AdS black hole. Under this condition, equations (\ref{TEQ}) and (\ref{FKERR}) simplify to
\begin{equation}\label{TEQJ}
    T(S,C)=\frac{\pi C +3S}{4\pi^{3/2}l\sqrt{CS}},\,\,\,\,\,\,\,F(S,C)=\frac{\pi \sqrt{C^3S}-\sqrt{CS^{3}}}{4\pi^{3/2}lC}.
\end{equation}
Examining this equation reveals that both $T$ and $F$ attain minima, denoted as $T_{\text{min}}$ and $F_{\text{min}}$, respectively, at $S=S_{\text{min}}$. These minima are expressed as
\begin{equation}\label{FTKERR}
    T_{min}=\frac{\sqrt{3}}{2\pi l} ,\,\,\,\,\,\,\,F_{min}=\frac{C}{6\sqrt{3}l},\,\,\,\,\,\,\, S_{min}=\frac{\pi C}{3}.
\end{equation}
Let us define the relative parameters as
\begin{equation}
    \Bar{s}=\frac{S}{S_{min}},\,\,\,\,\,\,\Bar{t}=\frac{T}{T_{min}},\,\,\,\,\,\Bar{f}=\frac{F}{F_{min}}.
\end{equation}
Substituting these relative parameters, the temperature and free energy equations (\ref{FTKERR}) can be reformulated as
\begin{equation}\label{bartf}
    \Bar{t}=\frac{\Bar{s}+1}{2\sqrt{\Bar{s}}},\,\,\,\,\,\,\,\Bar{f}=\frac{\sqrt{\Bar{s}}(3-\Bar{s})}{2}.
\end{equation}

\begin{figure}
    \centering
    \begin{subfigure}[b]{0.50\textwidth}
    \centering
    \includegraphics[width=\textwidth]{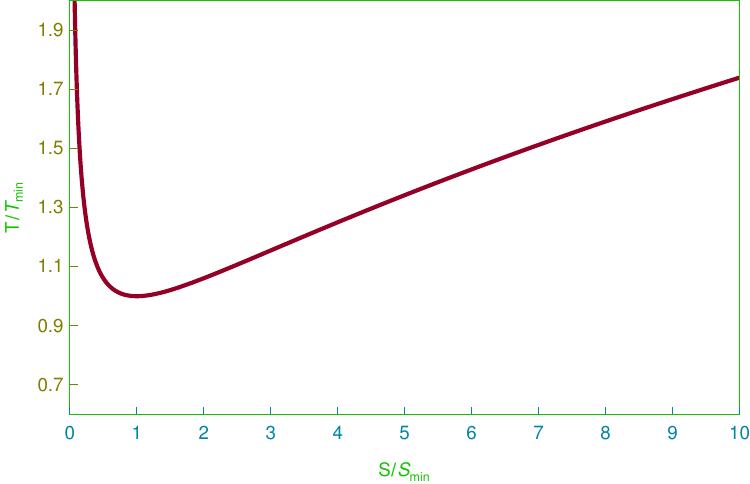}
         \caption{}
         \label{fig: TMINSMIN KERR}
     \end{subfigure}
     \hfill
     \begin{subfigure}[b]{0.45\textwidth}
    \centering
    \includegraphics[width=\textwidth]{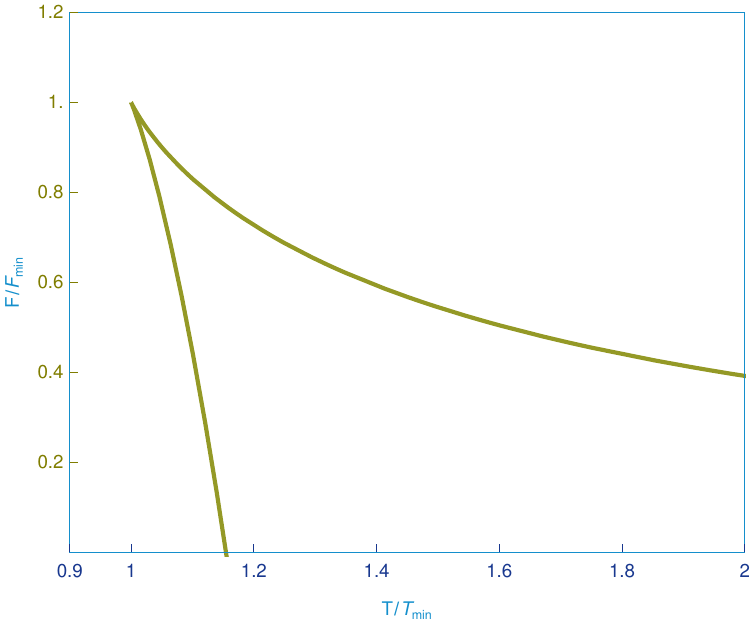}
    \caption{}
     \label{fig: FMINTMIN KERR}
     \end{subfigure}

     \caption{$T-S$ and $F-T$ curves at $J=0$}
     \label{fig: FT MIN KERR}
\end{figure}

The plots of $T-S$ and $F-T$ curves are shown in Fig. \ref{fig: FT MIN KERR}. From the $T-S$ curve we see that two black hole states correspond to the same $T, C$ but with different $S$ at $T>T_{min}$. The black hole state with large $S$, as well as a larger radius, is more stable than the lower entropy or smaller radius black hole. So there are jumps from a lower entropy state to a higher one can occur. However, this transition is not the first-order phase transition as there is no specific transition temperature. 

\subsubsection{\textit{T-S process at fixed} $\Omega, C$}
\begin{figure}
    \centering
    \begin{subfigure}[b]{0.50\textwidth}
    \centering
    \includegraphics[width=\textwidth]{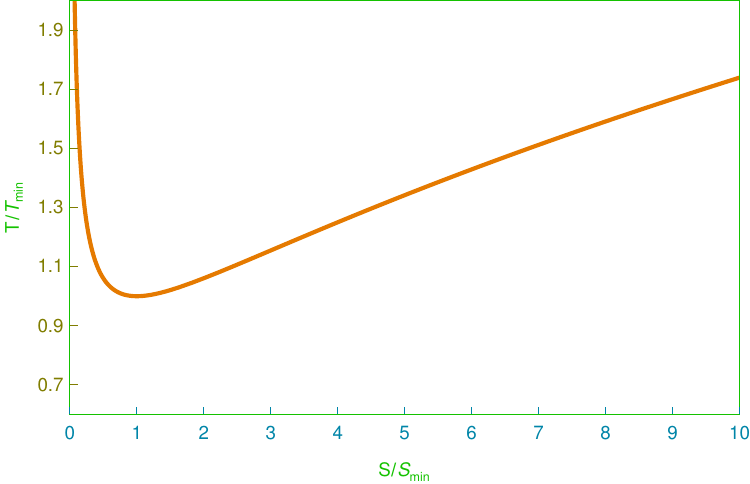}
         \caption{}
         \label{fig: TMINSMIN1 KERR}
     \end{subfigure}
     \hfill
     \begin{subfigure}[b]{0.45\textwidth}
    \centering
    \includegraphics[width=\textwidth]{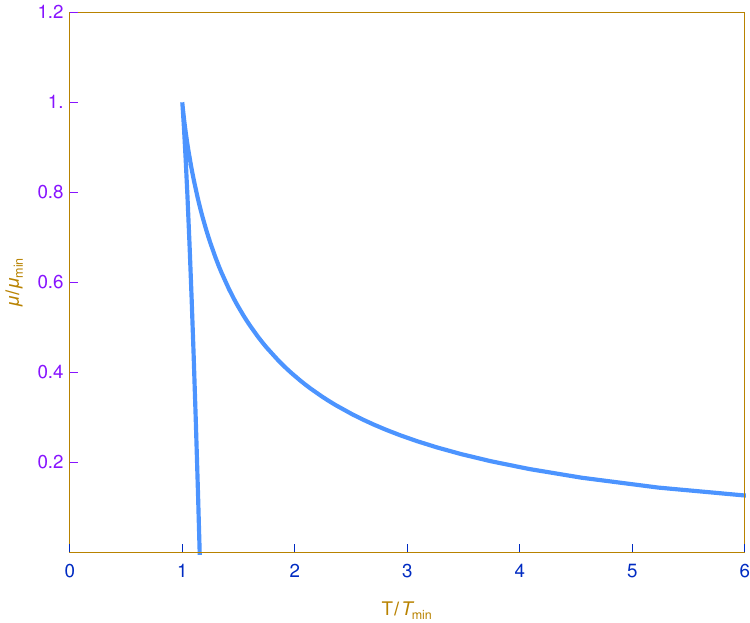}
    \caption{}
     \label{fig: mut KERR}
     \end{subfigure}

     \caption{$T-S$ and $\mu-T$ curves at fixed $\Omega$.}
     \label{fig: muts KERR}
\end{figure}

To analyze the $T-S$ process at fixed $\Omega$, we must express $T$ as a function of $S$, $C$, and $\Omega$. This can be achieved by substituting $J$ from equation (\ref{omegakerr}) into Equation (\ref{TEQ}). Thus, $T(S,C, \Omega)$ is given by
\begin{equation}\label{TOMEGA}
    T(S,C,\Omega)=\frac{2\pi C+6S-3Sl^2\Omega^2}{8\pi^{3/2}l\sqrt{CS}}.
\end{equation}
We observe that $T$ exhibits a minimum at $S=S_{\text{min}}$. These minima are expressed as
\begin{equation}\label{SMINKERR5}
    T_{min}=\frac{1}{2\pi l}\sqrt{\frac{3(2-l^2\Omega^2)}{2}},\,\,\,\,\,\,\,S_{min}=\frac{2\pi C}{3l(2-l^2\Omega^2)}.
\end{equation}
\textcolor{red}{E}quation (\ref{TOMEGA}) can be rescaled using the relative parameters\textcolor{red}{,} $\Bar{t}=\frac{T}{T_{\text{min}}}$ and $\Bar{s}=\frac{S}{S_{\text{min}}}$, resulting in
\begin{equation}\label{tplot1}
    \Bar{t}=\frac{\Bar{s}+1}{2\sqrt{\Bar{s}}}.
\end{equation}
This expression precisely matches the expression of $\Bar{t}$ in equation (\ref{bartf}). Therefore, the $T-S$ curve at fixed $\Omega$, depicted in Fig. \ref{fig: TMINSMIN1 KERR}, exhibits identical behavior to Fig. \ref{fig: TMINSMIN KERR}.

\subsubsection{$\mu-T$ \textit{process at fixed} $C,\Omega$}
Utilizing equations (\ref{muKERR}) and (\ref{omegakerr}), we can express the expression of $\mu$ in terms of $S$, $C$, and $\Omega$ as
\begin{equation}\label{mukerrf}
    \mu(S,C,\Omega)=\frac{\sqrt{S}[S(l^2\Omega^2-2)+2\pi C]}{8l\sqrt{\pi^3C^3}}.
\end{equation}
It is evident that at $S=S_{\text{min}}$, $\mu$ reaches its minimum value, denoted as $\mu_{\text{min}}$, which is expressed as
\begin{equation}
    \mu_{min}=\frac{1}{3l\sqrt{6(2-l^2\Omega^2)}}.
\end{equation}
Introducing the relative parameter $\Bar{\mu}=\frac{\mu}{\mu_{\text{min}}}$, we can rewrite equation (\ref{mukerrf}) as
\begin{equation}\label{muplot1}
    \Bar{\mu}=\frac{\sqrt{\Bar{s}}(3-\Bar{s})}{2}.
\end{equation}
Utilizing equations (\ref{muplot1}) and (\ref{tplot1}), we plot $\mu-T$ curves, as depicted in Fig. \ref{fig: mut KERR}. It's crucial to note the presence of a Hawking-Page transition occurring at a critical temperature $T_{\text{HP}}$. From equation (\ref{muplot1}), we observe that $\mu(T_{\text{HP}},\Omega)=0$ at $\Bar{s}=3$. Substituting this condition into equation (\ref{tplot1}), we obtain the Hawking-Page transition temperature, which is given by
\begin{equation}
    T_{HP}=\frac{2 T_{min}}{\sqrt{3}}.
\end{equation}

\begin{figure}
    \centering
     \centering
    \includegraphics[width=0.50\textwidth]{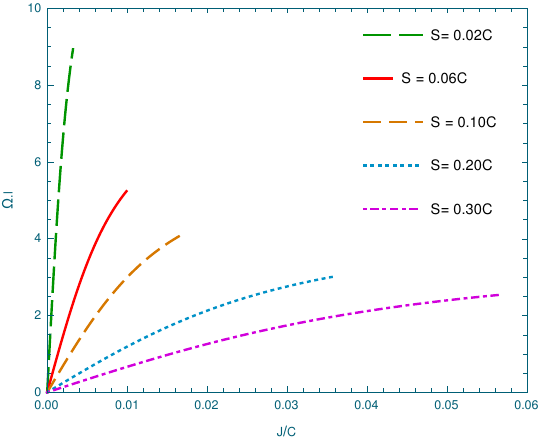}
      \caption{$\Omega-J$ curves at fixed $S$.}
     \label{fig: OMEGA J curve}
\end{figure}

\subsubsection{$\Omega-J$ \textit{process at fixed S, C}}
In the slow rotation limit, equation (\ref{omegakerr}) reveals a linear $\Omega-J$ relation, resulting in straight-line curves. However, in the non-limiting case described by equation (\ref{omegaKERR}), we discover a nontrivial relationship between $\Omega$ and $J$, offering deeper insights into the characteristics of the $\Omega-J$ process.

Exploiting the proportionality of $S$ and $J$ to $C$, we express the right-hand side of the equation (\ref{omegaKERR}) independently of $C$. Thus, equation (\ref{omegaKERR}) can be rewritten as
\begin{equation}
    \Omega=\frac{2\pi^{3/2}\mathcal{J}}{l}\sqrt{\frac{\mathcal{S+\pi}}{\mathcal{S}[4\pi^3\mathcal{J}^2+\mathcal{S}^2(\mathcal{S}+\pi)]}}.
\end{equation}
Where $\mathcal{J}=J/C$ and $\mathcal{S}=S/C$.

Referring to equation (\ref{TKERR2}), we observe the existence of an upper bound on $J$. Expressing this upper bound in terms of the rescaled variables $\mathcal{J}$ and $\mathcal{S}$, we find

\begin{equation}
 \mathcal{J}_{max}=\frac{\mathcal{S}\sqrt{3\mathcal{S}^2+4\pi \mathcal{S}+\pi^2}}{2\pi^2}.
\end{equation}
The endpoints of the $\Omega-J$ curves are determined by $\mathcal{J}_{max}$. 

The $\Omega-J$ curves at fixed $S$ are illustrated in Fig. \ref{fig: OMEGA J curve}. Notably, there is no indication of a phase transition; the curves originate from the origin and extend to $\mathcal{J}_{\text{max}}$. In the small $J$ range, the curves exhibit straight-line behavior, confirming the linear trend observed in slow rotating cases.

\begin{figure}
    \centering
     \centering
    \includegraphics[width=0.50\textwidth]{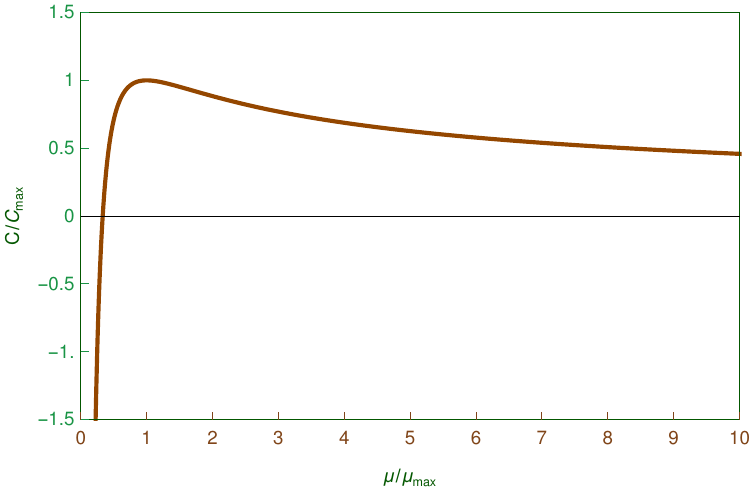}
      \caption{$\mu-C$ curve at fixed $(S,J)$}
     \label{fig: mu c kerr curve}
\end{figure}

\subsubsection{$\mu$-\textit{C process at fixed S, J}} 
To investigate the $\mu-C$ process, we examine the equation of state given by equation (\ref{muKERR}). It becomes evident that $\mu$ reaches a maximum value, denoted as $\mu=\mu_{\text{max}}$, at a certain $C=C_{\text{max}}$. These values are expressed as
\begin{equation}
    \mu_{max}=\frac{1}{6lS^3}\sqrt{\frac{(2\pi^2J^2+S^2)^3}{3}},\,\,\,\,\,\,\,\, C_{max}=\frac{3S^3}{\pi(2\pi^2J^2+S^2)}.
\end{equation}
Let us define the relative parameters as
\begin{equation}
    \Bar{\mu}=\frac{\mu}{\mu_{max}},\,\,\,\,\,\,\,\,c=\frac{C}{C_{max}}.
\end{equation}
In terms of these variables, equation (\ref{muKERR}) is expressed as
\begin{equation}
    \Bar{\mu}=\frac{3c-1}{2c^{3/2}}.
\end{equation}
The resemblance of this equation to equation (\ref{mucrn}) highlights the similarity in the $\mu-C$ process for both RN-AdS and Kerr black holes, suggesting the universality of this phenomenon. Figure \ref{fig: mu c kerr curve} illustrates the $\mu-C$ behavior at fixed $S$ and $J$.

\section{Thermodynamics of Kerr-Newman-AdS black hole at restricted phase space}

This section explores the critical behavior of a charged rotating black hole in AdS space. We shall first compute the first law, the Smarr formula, and the Gibbs-Duhem relation of the black hole thermodynamics. To do so we compute the conjugate to the thermodynamic variables. The metric of the KN-AdS black hole and the accompanying electromagnetic vector potential in the Boyer-Lindquist coordinates are
\cite{Caldarelli_1999}

\begin{equation}
   \begin{split}
    ds^2=-\frac{\Delta_r}{\rho^2}\left(dt-\frac{a \sin^2\theta}{\Sigma}d\phi\right)^2+&\frac{\rho^2}{\Delta_r}dr^2+\frac{\rho^2}{\Delta_\theta}d\theta^2+\frac{\sin^2\theta \Delta_\theta}{\rho^2}\left(adt-\frac{r^2+a^2}{\Sigma}d\phi^2\right)^2\\
    & A_\mu=-\frac{qr}{\rho^2}(\delta^t_\mu-\frac{a\sin^2\theta}{\Sigma}\delta^\phi_\mu).
   \end{split}
\end{equation}
Where
\begin{equation}
    \begin{split}
        &\rho^2=r^2+a^2\cos^2\theta,\,\,\,\,\,\,\Sigma=1-\frac{a^2}{l^2},\\
        &\Delta_r=(r^2+a^2)\left(1+\frac{r^2}{l^2}\right)-2mrG+Gq^2,\,\,\,\,\,\,\,\,\Delta_\theta=1-\frac{a^2}{l^2}\cos^2\theta.
    \end{split}
\end{equation}

The parameters $a$ and $q$ correspond to the rotation and electric charge of the black hole, respectively. The parameter $l$ corresponds to the AdS radius and is related to the cosmological constant by $\Lambda = -\frac{3}{l^2}$. Additionally, the restriction on $a$ and $l$ is given by $l^2 > a^2$.

The solution to $\Delta_r(r)=0$ yields two positive roots: a larger one, $r_+$, corresponding to the outer event horizon, and a smaller one, $r_-$, corresponding to the inner Cauchy horizon. By considering the equation $\Delta_r(r_+)=0$, $m$ can be expressed in terms of $r_+$, $a$, $l$, $q$, and $G$ as follows
\begin{equation}
    m=\frac{(r_+^2+a^2)(r_+^2+l^2)}{2Gl^2r_+}+\frac{q^2}{2r_+}
\end{equation}
The ADM mass, angular momentum, and charge can be expressed in terms of the quantities $m$, $a$, and $q$ as
\begin{equation}
    M=\frac{m}{\Sigma^2},\,\,\,\,\,\,\,\,J=\frac{am}{\Sigma^2},\,\,\,\,\,\,\,\,Q=\frac{q}{\Sigma}.
\end{equation}
Hence, the black hole mass can be expressed as
\begin{equation}\label{MKN}
    M=\frac{1}{2G\Sigma^2r_+}\left[r_+^2+a^2+\frac{r_+^4}{l^2}+\frac{a^2r_+^2}{l^2}+Gq^2\right].
\end{equation}
The other pair of conjugate thermodynamic quantities are given by,
\begin{itemize}
    \item the Bekenstein-Hawking entropy $S$, which is proportional to the area of the event horizon, and its conjugate, the Hawking temperature $T$, are
    \begin{equation}\label{STKN}
        S=\frac{\pi(r_+^2+a^2)}{G\Sigma},\,\,\,\,\,\,\,T=\frac{r_+}{4\pi(r_+^2+a^2)}\left(1+\frac{a^2}{l^2}+\frac{3r_+^2}{l^2}-\frac{a^2}{r_+^2}-\frac{Gq^2}{r_+^2}\right),
    \end{equation}
\end{itemize}
\begin{itemize}
    \item the angular momentum $J$ and its conjugate, the angular velocity at the event horizon $\Omega$, are given by
    \begin{equation}\label{OMEGAJKN}
        J=\frac{a}{2G\Sigma^2r_+}\left[r_+^2+a^2+\frac{r_+^4}{l^2}+\frac{a^2r_+^2}{l^2}+Gq^2\right],\,\,\,\,\,\,\Omega=\frac{a\Sigma}{r_+^2+a^2}+\frac{a}{l^2},
    \end{equation}
\end{itemize}
\begin{itemize}
    \item following the AdS-CFT dictionary \cite{PhysRevLett.130.181401}, the rescaled electric charge $\Bar{Q}$ and its conjugate, the electric potential $\Bar{\Phi}$ in the dual CFT, are given by 
    \begin{equation}\label{PHIQKN}
        \Bar{Q}=\frac{Ql}{\sqrt{G}},\,\,\,\,\,\,\,\,\Bar{\Phi}=\frac{\sqrt{G}qr_+}{l(r_+^2+a^2)}=\frac{\Phi \sqrt{G}}{l},
    \end{equation}
    here, $Q$ and $\Phi$ are defined in the bulk.
\end{itemize}
\begin{itemize}
    \item Unlike standard black hole thermodynamics in the RPS formalism, where the gravitational constant is treated as a constant, here it is considered as a variable. Correspondingly, in the dual CFT, the central charge $C$ is introduced as an extensive thermodynamic variable. This introduces an extra pair of conjugate variables: the central charge $C$ and its thermodynamic conjugate, the chemical potential $\mu$ \cite{Gao_2022}. These are given by
    \begin{equation}\label{KNCmu}
        C=\frac{l^2}{G},\,\,\,\,\,\,\,\,\,\,\,\mu=\frac{M-TS-\Omega J-\Bar{\Phi}\Bar{Q}}{C}.
    \end{equation}
While we introduced $\mu-C$ following the AdS/CFT duality description, it's crucial to remember that $\mu$ is defined as the chemical potential in the bulk, and $C$ is defined as the number of microscopic degrees of freedom in the bulk, i.e., $\mu_{\text{CFT}}=\mu_{\text{bulk}}$ and $C=N_{\text{bulk}}$. Our study focuses on the phase space of thermodynamic variables in the bulk rather than in the dual CFT. Additionally, it's worth mentioning that, unlike Visser's formalism, where variations of the central charge are considered, we vary the gravitational constant but not the AdS radius.
\end{itemize}
It is straightforward to verify that all the thermodynamic variables mentioned here satisfy the first law of the RPS formalism,
\begin{equation}\label{firstlawKN}
    dM=TdS+\Bar{\Phi}d\Bar{Q}+\Omega dJ+\mu dC.
\end{equation}
From equation (\ref{KNCmu}), we observe that the Euler relation holds, similar to standard thermodynamics,
\begin{equation}\label{EulerKN}
    M=TS+\Bar{\Phi}\Bar{Q}+\Omega J+\mu C.
\end{equation}
We rescale entropy, electric charge, and angular momentum by the central charge as follows
\begin{equation}
    \mathcal{S}=\frac{S}{C},\,\,\,\,\,\,\,\mathcal{\Bar{Q}}=\frac{\Bar{Q}}{C},\,\,\,\,\,\,\,\,\,\mathcal{J}=\frac{J}{C}.
\end{equation}
Applying this rescaling to equations (\ref{firstlawKN}) and (\ref{EulerKN}), we obtain the relation,
\begin{equation}
    d\mu=-\mathcal{S}dT-\mathcal{\Bar{Q}}d\Bar{\Phi}-\mathcal{J}d\Omega.
\end{equation}
This is known as the Gibbs-Duhem relation, which is a distinctive feature of the RPS formalism \cite{Gao_2022}. Importantly, it is absent in both standard black hole thermodynamics and EPST.

\subsection{Equation of states}

To initiate the thermodynamic analysis, it is essential to express the mass of the black hole $M$ and the intensive variables $T, \Bar{\Phi}, \Omega$, and $J$ in terms of the extensive variables $S, \Bar{Q}$, and $C$. To achieve this, we need to express $a$ and $G$ in terms of $J$, $M$, and $C$\textcolor{red}{,}
\begin{equation}\label{GaKN}
    G=\frac{l^2}{C},\,\,\,\,\,\,\,\,\,\,a=\frac{J}{M}.
\end{equation}
Utilizing equations (\ref{STKN}) and (\ref{GaKN}), we can express the expression of $r_+$ as
\begin{equation}\label{rKN}
    r_+^2=\frac{S(M^2l^2-J^2)}{\pi C M^2}-\frac{J^2}{M^2}.
\end{equation}
Substituting equations (\ref{GaKN}) and (\ref{rKN}) into (\ref{MKN}), we can explicitly express the mass in terms of extensive variables,
\begin{equation}\label{MKN1}
    M(J,C,S,\Bar{Q})=\frac{[4J^2C\pi^3(C\pi+S)+(\pi^2\Bar{Q}^2+\pi C S+S^2)^2]^{1/2}}{2l\pi^3/2\sqrt{CS}}.
\end{equation}
Utilizing equations (\ref{MKN1}) and (\ref{firstlawKN}), we obtain the equation of state as follows,
\begin{equation}\label{TKN1}
    T=\left(\frac{\partial M}{\partial S}\right)_{J,C,\Bar{Q}}=\frac{3S^4+4\pi C S^3+\pi S^2(C^2+2\Bar{Q}^2)-\pi^4(4C^2J^2+\Bar{Q}^4)}{4l\sqrt{C}\pi^{3/2}S^{3/2}[{4J^2C\pi^3(C\pi+S)+(\pi^2\Bar{Q}^2+C\pi S+S^2)^2}]^{1/2}},
\end{equation}
\begin{equation}\label{OMEGAKN1}
    \Omega=\left(\frac{\partial M}{\partial J}\right)_{S,\Bar{Q},C}=\frac{2J\pi^{3/2}\sqrt{C}(\pi C+S)}{l\sqrt{S}[{4J^2C\pi^3(C\pi+S)+(\pi^2\Bar{Q}^2+C\pi S+S^2)^2}]^{1/2}},
\end{equation}
\begin{equation}\label{PHIKN1}
    \Bar{\Phi}=\left(\frac{\partial M}{\partial \Bar{Q}}\right)_{S,J,C}=\frac{\sqrt{\pi}\Bar{Q}(\pi^2\Bar{Q}^2+\pi C S+S^2)}{l\sqrt{CS}[{4J^2C\pi^3(C\pi+S)+(\pi^2\Bar{Q}^2+C\pi S+S^2)^2}]^{1/2}},
\end{equation}
\begin{equation}\label{muKN1}
    \mu=\left(\frac{\partial M}{\partial C}\right)_{J,\Bar{Q},S}=\frac{\pi^4(4J^2C^2-\Bar{Q}^4)+\pi^2S^2(C^2-2\Bar{Q^2})-S^4}{4l\pi^{3/2}C^{3/2}\sqrt{S}[{4J^2C\pi^3(C\pi+S)+(\pi^2\Bar{Q}^2+C\pi S+S^2)^2}]^{1/2}}.
\end{equation}
It can be readily verified that the temperature, angular momentum, electric potential, and chemical potential described in the equation of states (\ref{TKN1})-(\ref{muKN1}) coincide with the relations (\ref{STKN})-(\ref{KNCmu}). Furthermore, the aforementioned results align with the RN-AdS equation of state when $J=0$ and with the Kerr equation of state when $\Bar{Q}=0$.

Now, we can define an interesting quantity: the heat capacity at a constant electric charge, angular momentum, and central charge, denoted as $\mathcal{C}|_{\Bar{Q}, J, C}$. Utilizing equation (\ref{TKN1}), we can express it as
\begin{equation}
    \begin{split}
        \mathcal{C}|_{\Bar{Q}, J, C}&=T\left(\frac{\partial S}{\partial T}\right)_{\Bar{Q}, J, C}\\&=(2S(-\pi^4(4J^2C^2+\Bar{Q}^4)+\pi^2(P^2+2\Bar{Q}^2)S^2+4C\pi S^3+3S^4)(4J^2C\pi^3(C\pi+S)\\+&(\pi^2\Bar{Q}^2+C\pi C+S^2)^2))/(16J^4C^3\pi^7(3C\pi+4S)+(\pi^2\Bar{Q}^2+C\pi S+S^2)^3\\&(3\pi^2\Bar{Q}^2-C\pi S+3S^2)+8J^2C\pi^3(3C\pi^3(3C\pi^5\Bar{Q}^4+2\pi^4\Bar{Q}^2(2C^2+\Bar{Q}^2)S\\&+3C\pi^3(C^2+2\Bar{Q}^2)S^2+12C^2\pi^2S^3+15C\pi^2S^4+6S^5)).
    \end{split}
\end{equation}
The positivity of $\mathcal{C}|_{\Bar{Q}, J, C}$ ensures the stability of the thermodynamic system under small fluctuations of the thermodynamic variables in the canonical ensemble \cite{Cvetic_1999}. This implies that the critical hypersurfaces of phase space can be obtained by examining the divergence of $\mathcal{C}|_{\Bar{Q}, J, C}$. Such surfaces are given by
\begin{equation}
    \begin{split}
        \mathcal{J}^2=&-\frac{1}{4\pi^4(3\pi+4\mathcal{S})}(3\pi^5\mathcal{\Bar{Q}}^4+2\pi^4\mathcal{\Bar{Q}}^2(2+\mathcal{\Bar{Q}}^2)\mathcal{S}+3\pi^3(1+2\mathcal{\Bar{Q}}^2)\mathcal{S}^2+12\pi^2\mathcal{S}^3\\&+15\pi \mathcal{S}^4+6\mathcal{S}^5-2[\mathcal{S}^2(\pi^8\mathcal{\Bar{Q}}^4(-2+\mathcal{\Bar{Q}}^2)^2+3\pi^7\mathcal{\Bar{Q}}^2(2+3\mathcal{\Bar{Q}}^2-2\mathcal{\Bar{Q}}^4)\mathcal{S}\\&+3\pi^6)(1+8\mathcal{\Bar{Q}}^2+2\mathcal{\Bar{Q}}^4)\mathcal{S}^2+\pi^5(19+6\mathcal{\Bar{Q}}^2(6+\mathcal{\Bar{Q}}^2))\mathcal{S}^3+6\pi^4(9+4\mathcal{\Bar{Q}}^2+\mathcal{\Bar{Q}}^4)\mathcal{S}^4\\&+3\pi^3(29+2\mathcal{\Bar{Q}}^2)\mathcal{S}^5+82\pi^2\mathcal{S}^6+42\pi\mathcal{S}^7+9\mathcal{S}8]^{1/2})).
    \end{split}
\end{equation}
Here, we have employed the scaling $S=\mathcal{S}C$, $J=\mathcal{J}C$, and $\Bar{Q}=\mathcal{\Bar{Q}}C$. 

\textit{Homogeneity:} To verify homogeneity, we rescaled $S$, $\Bar{Q}$, $J$, and $C$ as $S\rightarrow \alpha S$, $\Bar{Q}\rightarrow \alpha \Bar{Q}$, $J\rightarrow \alpha J$, and $C\rightarrow \alpha C$ respectively. From equation (\ref{MKN1}), we observe that $M$ scales as $\alpha M$, while equations (\ref{TKN1})-(\ref{muKN1}) show that $T$, $\Omega$, $\Bar{\Phi}$, and $\mu$ remain unaffected by this rescaling. This confirms that $M$ is a first-order homogeneous function, and $T$, $\Omega$, $\Phi$, and $\mu$ are zeroth-order homogeneous functions. This zeroth-order homogeneity verifies the intensive character of $T$, $\Omega$, $\Phi$, and $\mu$.

From a simple analysis, it can be shown that $S$, $\Bar{Q}$, and $J$ are proportional to the central charge. As we mentioned, $C$ plays the role of the number of particles, so it is extensive by nature. The proportionality of $S$, $\Bar{Q}$, and $J$ with $C$ establishes their extensivity.

The positive definiteness of $T$ implies an upper bound on $J$. From equation (\ref{TKN1}), we find that the upper bound can be expressed in terms of macrostate variables as
\begin{equation}\label{LIMITOFJKN}
    J\leq J_{max}=\frac{[C^2\pi^2 S^2+2\pi^2\Bar{Q}^2S^2+4C\pi S^3+3S^4-\pi^4\Bar{Q}^4]^{1/2}}{2C\pi^2}.
\end{equation}

\subsection{Thermodynamic process and critical behavior}
With four constraint equations (\ref{TKN1})-(\ref{muKN1}) and eight thermodynamic variables $T, J,\Omega, C$, $S,\Bar{Q},\Bar{\Phi}$, and $\mu$, the thermodynamic phase space of the KN-AdS black hole is characterized by four variables. From the constraint equations, we observe that each intensive variable is a function of four extensive variables. To simplify the process, we vary one extensive variable while keeping others constant. This leads to various macroscopic processes. In the subsequent sections, we focus on a few of these processes and analyze the critical behavior of the corresponding curves in the thermodynamic phase space.

 \subsubsection{\textit{Critical point}}
In this section, we identify the critical points that will aid in studying the critical behavior of the KN-AdS black hole in a later section. As discussed in Sec [\ref{TS PROCESS RN}] and Sec [\ref{TS PROCESS KERR}], we have already observed first-order and second-order phase transitions at critical points in the $T-S$ curves of the RN and Kerr black holes. 

The critical points for the Van der Waals-like phase transition on the $T-S$ curves of the KN-AdS black hole are obtained by following these restriction equations,
 \begin{equation}\label{INFLECTION}
      \left(\frac{\partial T}{\partial S}\right)_{J,C,\Bar{Q}}=0,\,\,\,\,\,\,\,\,\, \left(\frac{\partial^2 T}{\partial S^2}\right)_{J,C,\Bar{Q}}=0.
 \end{equation}
Utilizing equation (\ref{TKN1}), equations (\ref{INFLECTION}) are reduced to
\begin{equation}\label{CRITICALKN1}
    \begin{split}
    &16J^4C^3\pi^7(3C\pi+4S)+(\pi^2\Bar{Q}^2+C\pi S+S^2)^3(3\pi^2\Bar{Q}^2-C\pi S+3S^2)+8J^2C\pi^3(3C\pi^5\Bar{Q}^4\\&+2\pi^4\Bar{Q}^2(2C^2+\Bar{Q}^2)S+3C\pi^3(C^2+2\Bar{Q}^2)S^2+12C^2\pi^2S^3+15C\pi S^4+6S^5)=0,
    \end{split}
\end{equation}
\begin{equation}\label{CRITICALKN2}
\begin{split}
     &-(5\pi^2\Bar{Q}^2-C\pi S+S^2)(\pi^2\Bar{Q}^2+C\pi S+S^2)^5-64J^6C^4\pi^{10}(5C^2\pi^2+12C\pi S+8S^2)\\&-16J^4C^2\pi^6(15C^2\pi^6\Bar{Q}^4+24C\pi^5\Bar{Q}^2(C^2+\Bar{Q}^2)S+\pi^4(13C^4+58C^2\Bar{Q}^2+8\Bar{Q}^4)S^2\\&+40C\pi^3(C^2+\Bar{Q}^2)S^3+35C^2\pi^2S^4-8S^6)-4J^2C\pi^3(\pi^2\Bar{Q}^2+C\pi S+S^2)(15C\pi^7\Bar{Q}^6\\&+3\pi^6\Bar{Q}^4(11C^2+4\Bar{Q}^2)S+C\pi^5\Bar{Q}^2(25C^2+57\Bar{Q}^2)S^2+\pi^4(15C^4+56C^2\Bar{Q}^2+28\Bar{Q}^4)S^3)\\&+5C\pi^3(13C^2+\Bar{Q}^2)S^4+7\pi^2(15C^2-4\Bar{Q}^2)S^5+75C\pi S^6+20S^7)=0.
\end{split}
   \end{equation}
If we substitute $J=0$ or $\Bar{Q}=0$ into these equations, we retrieve the corresponding equations for the RN-AdS and Kerr-AdS black holes, respectively.

The solution of the above two equations yields critical points. In earlier sections, we observed that for the RN-AdS and Kerr-AdS cases at the slow-rotating limit, it is possible to find critical points analytically. However, for the KN-AdS case, it is impossible to solve equations (\ref{CRITICALKN1}) and (\ref{CRITICALKN2}) analytically. Therefore, we solve them numerically, and through fitting these numerical results, we obtain expressions for the critical parameters. 

\begin{figure}
    \centering
    \begin{subfigure}[b]{0.48\textwidth}
    \centering
    \includegraphics[width=\textwidth]{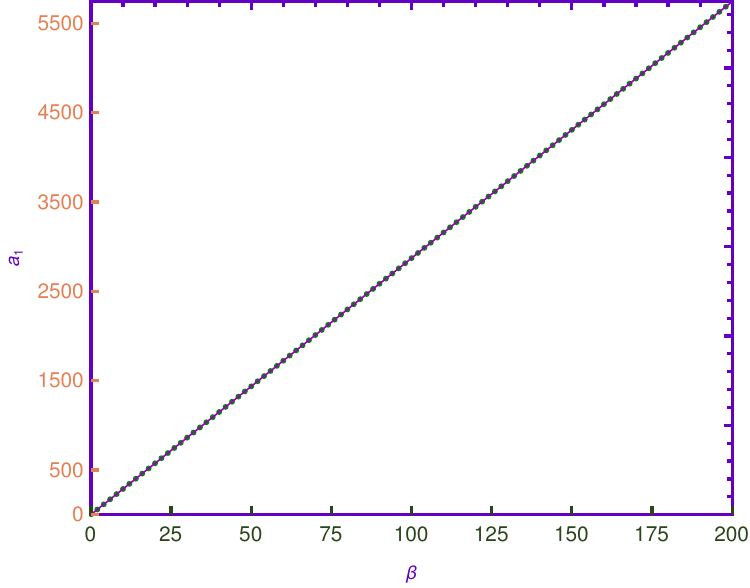}
         \caption{}
         \label{fig: a1vsbeta}
     \end{subfigure}
     \hfill
     \begin{subfigure}[b]{0.48\textwidth}
    \centering
    \includegraphics[width=\textwidth]{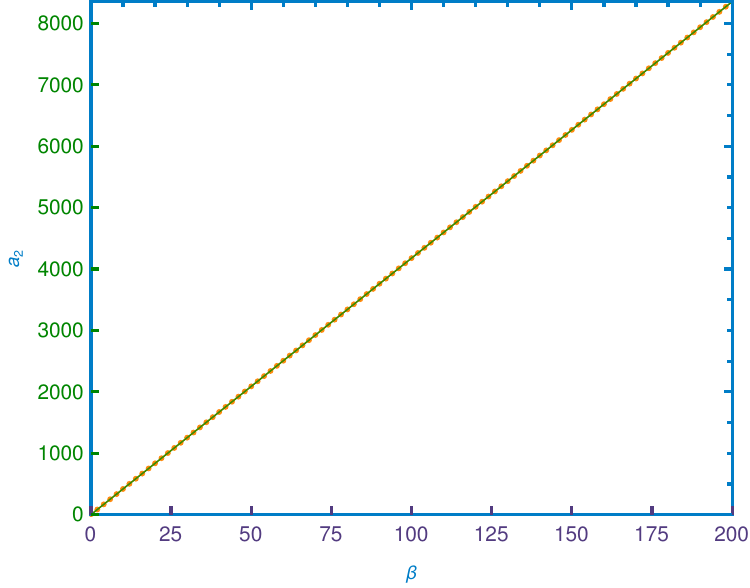}
    \caption{ }
     \label{fig: a2vsbeta}
     \end{subfigure}

     \caption{The behaviors of the coefficients $\alpha_1$ and $\alpha_2$ as functions of $\beta$ are displayed in Figures (a) and (b), respectively. The dots represent the exact numerical results, while the solid lines show the fitting results with the forms given in equations (\ref{alpha1}) and (\ref{alpha2}). The fitting is highly consistent.}
     \label{fig: a1a2vs beta}
\end{figure}

The parameters that describe the equation of states can be classified as universal parameters and characteristic parameters. Universal parameters represent universal properties of a thermodynamic system, like, pressure, volume, temperature, etc. Whereas the characteristic parameters describe characteristic properties of the system, like, critical pressure, critical temperature, etc \cite{PhysRevD.93.084015}. The RN-AdS and Kerr-AdS are systems of a single characteristic parameter. Whereas a KN-AdS black hole is a thermodynamic system described by two characteristic parameters, $\Bar{Q}\,\text{and}\, J$ \cite{Cheng_2016}. Note that there are four thermodynamic variables and two constraint equations (\ref{CRITICALKN1}) and (\ref{CRITICALKN2}). Therefore, all critical parameters can be expressed through two characteristic parameters. We define a dimensionless parameter $\beta$ as the ratio of angular momentum to rescaled electric charge,
\begin{equation}
    \beta=\frac{J}{\Bar{Q}}
\end{equation}.
Therefore, instead of $(J,\Bar{Q})$, we express the critical point in terms of $(\beta,\Bar{Q})$. Through dimensional analysis, these critical points can be written as
\begin{equation}\label{coefficient}
    S_c=a_1(\beta)\Bar{Q},\,\,\,\,\,\,\,\,C=a_2(\beta)\Bar{Q}.
\end{equation}
After substituting $J$, $S$, and $C$, equations (\ref{CRITICALKN1}) and (\ref{CRITICALKN2}) are reduced to equations of $a_1(\beta)$ and $a_2(\beta)$. We solve them numerically and find a fitting form very accurately, allowing us to express the coefficients $a_1$ and $b_1$ as functions of $\beta$. We find a fitting form with fewer unknown parameters compared to other works in this direction \cite{PhysRevD.108.044045, Cheng_2016}. The use of a large number of unknown parameters reduces the credibility of the fitting form due to the high AIC and BIC. In Fig. \ref{fig: a1a2vs beta}, the behaviors of $a_1$ and $a_2$ are depicted. Exact numerical results are represented by dots, and the solid lines correspond to the fitting forms given by equations (\ref{CRITICALKN1}) and (\ref{CRITICALKN2}).

The coefficient functions in equation (\ref{coefficient}) are given by
\begin{equation}\label{alpha1}
    a_1(\beta)=\frac{34.6618\beta^3-199.0466\beta^2-8.2662\beta-0.9354}{1.2069\beta^2-6.9309\beta-0.2935},
\end{equation}
\begin{equation}\label{alpha2}
    a_2(\beta)=\frac{171.2387\beta^3-1370.7869\beta^2-88.2334\beta-13.0112}{4.0982\beta^2-32.8068\beta-2.1506}.
\end{equation}

Substituting the critical parameters into equation (\ref{TKN1}), we obtain the critical temperature as
\begin{equation}\label{TCKN1}
    T_c=\frac{3a_1^4+4\pi a_1^3a_2+\pi^2a_1^2(a_2^2+2)-\pi^4(4\beta^2a_2^2+1)}{4l\pi^{3/2}a_1^{3/2}a_2^{1/2}[(a_1^2+\pi a_1a_2+\pi^2)^2+4\pi^3\beta^2a_2(a_1+\pi a_2)]^{1/2}}.
\end{equation}

Now, we can verify whether our critical parameters satisfy the critical parameters of RN-AdS and Kerr-AdS in the appropriate limits.

The value of the critical parameters at $\beta\rightarrow 0 $  $(J\approx 0)$ are:
\begin{equation}
    S_c=0.526785C,\,\,\,\,Q_c=0.165289C,\,\,\,\,\,T_c=0.260428/l.
\end{equation}

Comparing with equation (\ref{criticalrn}), we observe that the results match very well with the RN-AdS values. Therefore, we conclude that the numerical treatment and our fitting forms are very reliable for the rest of the analysis.

\begin{figure}
    \centering
    \begin{subfigure}[b]{0.45\textwidth}
    \centering
    \includegraphics[width=\textwidth]{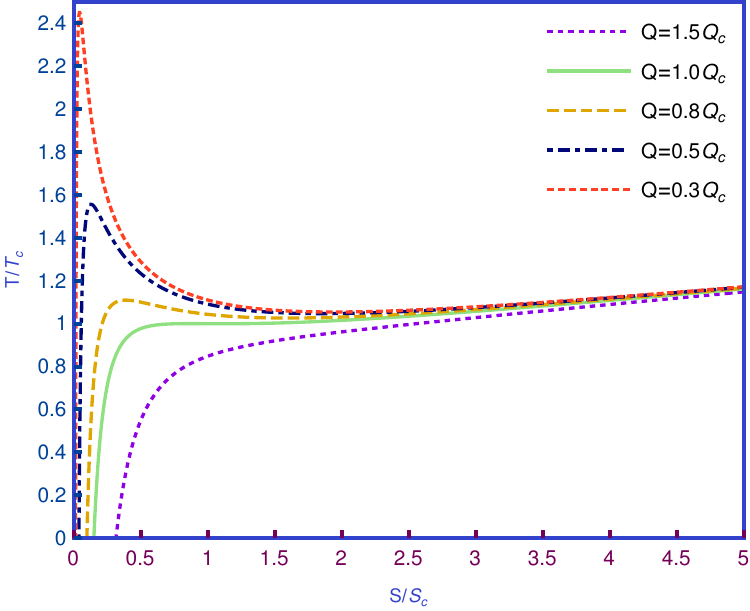}
         \caption{}
         \label{fig: KN for RN}
     \end{subfigure}
     \hfill
     \begin{subfigure}[b]{0.45\textwidth}
    \centering
    \includegraphics[width=\textwidth]{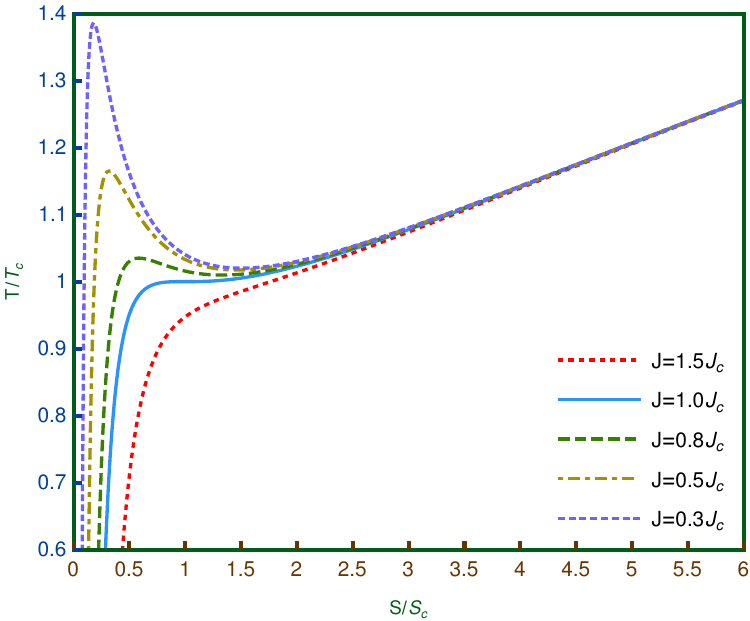}
    \caption{}
     \label{fig: KN for KERR }
     \end{subfigure}
     \hfill
    \begin{subfigure}[b]{0.45\textwidth}
    \centering
    \includegraphics[width=\textwidth]{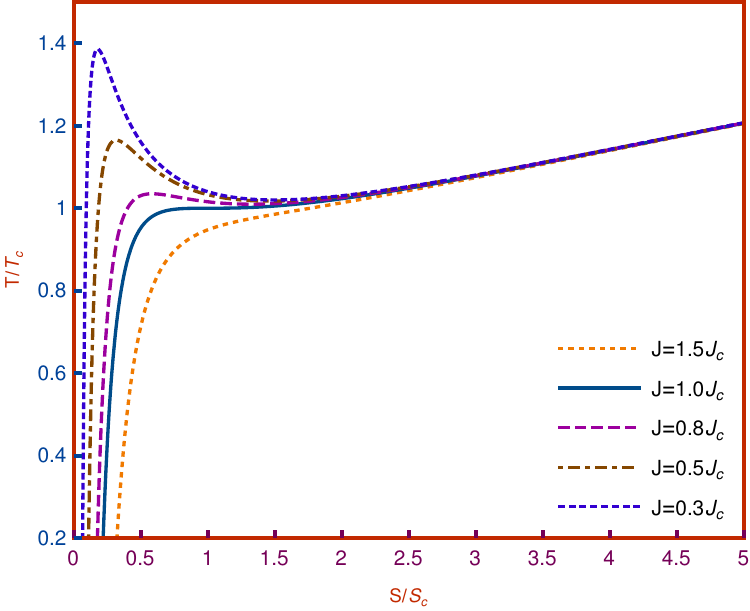}
    \caption{}
     \label{fig: KN ADS}
     \end{subfigure} 
    
     \caption{The plots depict $T-S$ curves. Figures (a), (b), and (c) illustrate the $T-S$ behavior using equation (\ref{tKN}) for $(J=0, \Bar{Q}\neq0)$, $(J\neq0, \Bar{Q}=0)$, and $(J\neq0, \Bar{Q}\neq0)$, respectively. Figures (a) and (b) represent the $T(S)$ curves of the RN-AdS and Kerr-AdS cases, respectively. Figure (c) illustrates the behavior of $T(S)$ curves of the KN-AdS black hole.}
     \label{fig: KN KERR RN}
\end{figure}

The values of the critical parameters as $\beta\rightarrow \infty$ $(\Bar{Q}\approx 0)$ are:
\begin{equation}
    S_c=0.687325C,\,\,\,\,\,\,J_c=0.023932C,\,\,\,\,\,\,\,\,T_c=0.269866/l.
\end{equation}

In comparison with (\ref{criticalvalues}), it is evident that at $Q=0$, we recover the critical parameters corresponding to the Kerr-AdS black hole.

\subsubsection{\texorpdfstring{\textit{T-S process at fixed} \(J,\Bar{Q},C\)}{T-S process at fixed J,\Bar{Q},C}} 

Now, we define the relative parameters as,
\begin{equation}
    s=\frac{S}{S_c}, \,\,\,\,\,\,\,\,\,\,\,\,t=\frac{T}{T_c},\,\,\,\,\,\,\,\,j=\frac{J}{J_c},\,\,\,\,\,\,q=\frac{Q}{Q_c}.
\end{equation}

Utilizing equations (\ref{TKN1}) and (\ref{TCKN1}), we find the expression of the relative temperature as

\begin{equation}\label{tKN}
\begin{split}
     t= &\frac{\sqrt{(a_1^2+a_1a_2\pi+\pi^2)^2+4a_2\pi^3(a_1+a_2\pi)\beta^2}}{\sqrt{(j^2\pi^2+a_1s(a_2\pi+a_1s))^2+4j^2n\pi^3(a_2\pi+a_1s)\beta^2}}\\&\times \frac{(-j^4\pi^4+a_1^2s^2(a_2\pi+a_1s)(a_2\pi+3a_1s)+2j^2(a_1^2\pi^2s^2-2a_2^2\pi^4\beta^2))}{s^{3/2}{(3a_1^4+4a_1^3a_2\pi +a_1^2(2+a_2^2)\pi^2-\pi^4(1+4a_2^2\beta^2))}}.
\end{split}
\end{equation}

Due to parameter scaling, the expression of the relative temperature is independent of the central charge $C$. This is known as the law of corresponding states.

In Figure \ref{fig: KN KERR RN}, we present the $T-S$ curves at fixed $J$, $\Bar{Q}$, and $C$. Here, we analyze the KN-AdS case and two limiting cases: one where $\beta\rightarrow0$, leading to the RN-AdS case \ref{fig: KN for RN}, and the other where $\beta\rightarrow\infty$, leading to the Kerr-AdS case \ref{fig: KN for KERR }.

In Fig. \ref{fig: KN ADS}, the general KN-AdS case is depicted where both $J$ and $\Bar{Q}$ are nonzero. Here, we observe that for $J<J_c$, there are non-monotonic behaviors of the $T(S)$ curve. There are two branches of stable black holes, one consisting of small black holes and the other of larger black holes. These two branches are separated by unstable black holes of intermediate sizes.

As $J<J_c$, the height of the local maxima of the $T-S$ curves increases with the decrease of $J$, and the curves become more non-monotonic. Thus, there are two black hole phases separated by an unstable phase. This behavior signals a van der Waals-like first-order phase transition in the range $0<J<J_c$.

At $J=J_c$, the local maxima become degenerated, and the first-order phase transition shifts to second-order.

For $J>J_c$, the behavior of the $T(S)$ curves changes. They become monotonic and exhibit a single black hole phase.

We observe such phenomena in EPST formalism, but they do not appear in standard black hole thermodynamics.

\subsubsection{\texorpdfstring{\textit{F-T process at fixed} \(J,\Bar{Q},C\)}{F-T process at fixed J,\Bar{Q},C}} 

Now, we introduce the Helmholtz free energy, which is the thermodynamic potential in the canonical ensemble.
\begin{equation}
    F(T,\Bar{Q},J,C)=M(T,\Bar{Q},J,C)-TS.
\end{equation}
Utilizing equations (\ref{MKN1}) and (\ref{TKN1}), the expression of $F$ can be written as
\begin{equation}\label{FKN1}
    F(S,J,\Bar{Q},C)=\frac{4J^2C\pi^3(3\pi C+2S)+(3\pi^2\Bar{Q}^2+C\pi S-S^2)(\pi^2\Bar{Q}^2+C\pi S+S^2)}{4l\pi^{3/2}\sqrt{CS}[4J^2C\pi^3(C\pi+S)+(\pi^2\Bar{Q}^2+C\pi S+S^2)^2]^{1/2}}.
\end{equation}
Introducing the critical parameters, we obtain the critical value of the Helmholtz energy,
\begin{equation}
    F_c=\frac{C(-a_1^4+a_1^2(2+a_2^2)\pi^2+4a_1a_2\pi^3(1+2\beta^2)+3\pi^4(1+4a_2^2\beta^2))}{4la_2\sqrt{a_1a_2}\pi^{3/2}[(a_1^2+a_1a_2\pi+\pi^2)^2+4a_2\pi^3(a_1+a_2\pi)\beta^2]^{1/2}}.
\end{equation}
Substituting $\beta\rightarrow 0$ and $\beta\rightarrow\infty$ into the above equation, we find that the values of $F_c$ in the RN-AdS and Kerr limits are $0.135245C/l$ and $0.105095$, respectively.

Comparing with equations (\ref{fcrnlimit}) and (\ref{fckerrlimit}), we observe that our limiting values agree well with the exact values.

\begin{figure}
    \centering
    \begin{subfigure}[b]{0.45\textwidth}
    \centering
    \includegraphics[width=\textwidth]{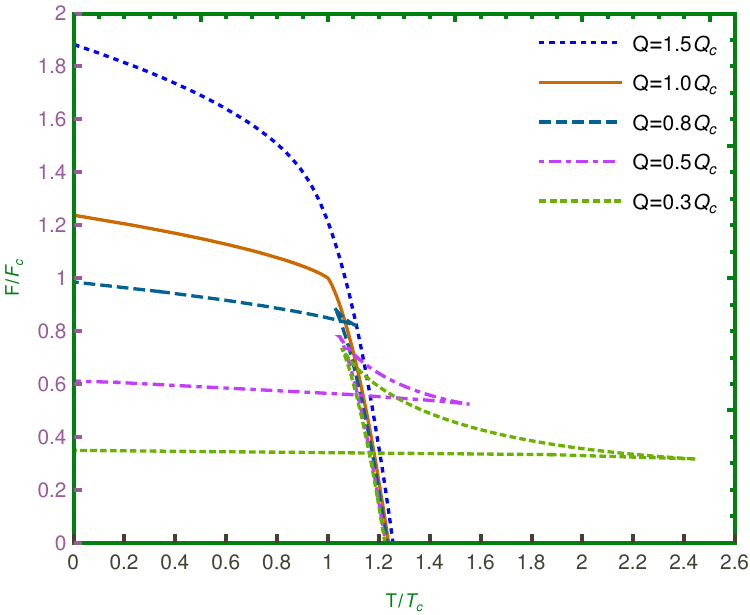}
         \caption{}
         \label{fig: FT PLOT KNRN}
     \end{subfigure}
     \hfill
     \begin{subfigure}[b]{0.45\textwidth}
    \centering
    \includegraphics[width=\textwidth]{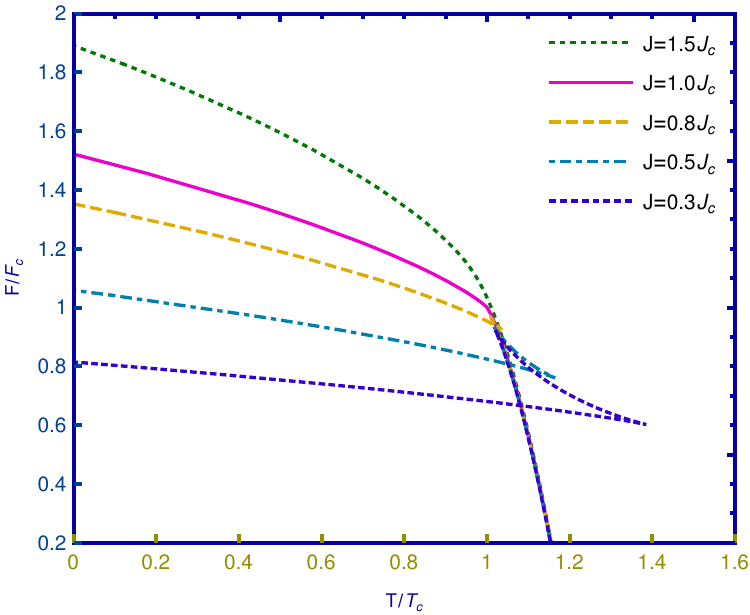}
    \caption{}
     \label{fig: FT PLOT KNKERR }
     \end{subfigure}
     \hfill
    \begin{subfigure}[b]{0.45\textwidth}
    \centering
    \includegraphics[width=\textwidth]{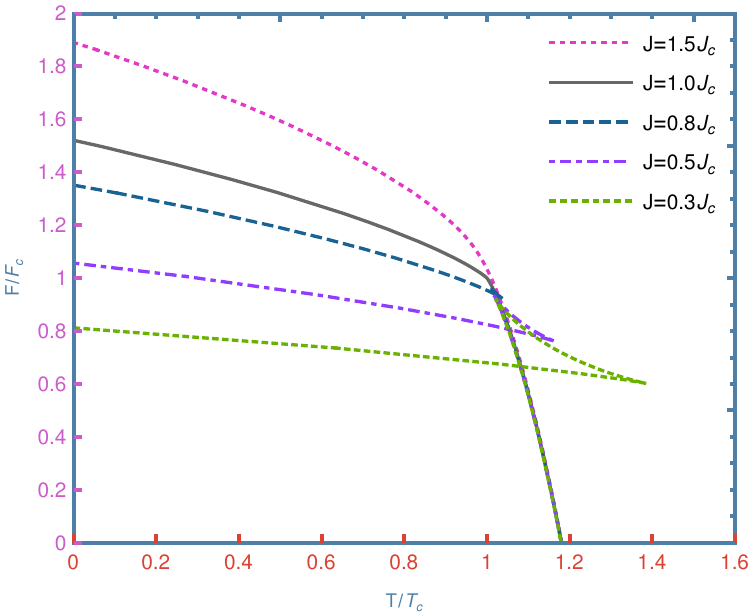}
    \caption{}
     \label{fig: FT PLOT KN}
     \end{subfigure} 
    
     \caption{ The plots depict $F-T$ curves. Figures (a), (b), and (c) illustrate the $F-T$ behavior using equations (\ref{tKN}) and (\ref{fKN}) for $(J=0, \Bar{Q}\neq0)$, $(J\neq0, \Bar{Q}=0)$, and $(J\neq0, \Bar{Q}\neq0)$, respectively. Figures (a) and (b) represent the $F(T)$ curves of the RN-AdS and Kerr-AdS cases, respectively. Figure (c) illustrates the behavior of $F(T)$ curves of the KN-AdS black hole.}
     \label{fig: FT PLOT KN RN KERR}
\end{figure}

Now, let's define the relative parameter as
\begin{equation}
    f=\frac{F}{F_c},
\end{equation}
\textcolor{red}{t}hen,
\begin{equation}\label{fKN}
    \begin{split}
        f=&\frac{\sqrt{(a_1^2+a_1a_2\pi +\pi^2)^2+4a_2\pi^3(a_1+a_2\pi)\beta^2}}{\sqrt{s((j^2\pi^2+a_1s(a_2\pi+a_1s))^2+4j^2a_2\pi^3(a_2\pi+a_1s)\beta^2)}}\\
        &\times\frac{(3j^4\pi^4+a_1^2a_2^2\pi^2s^2-a_1^4s^4+2j^2\pi^2(a_1s(2a_2\pi+a_1s)+2a_2\pi(3a_2\pi+2a_1s)\beta^2))}{(-a_1^4+a_1^2(2+a_2^2)\pi^2+4a_1a_2\pi^3(1+2\beta^2)3\pi^4(1+4a_2^2\beta^2))}.
    \end{split}
\end{equation}
Using the temperature equations (\ref{tKN}) and free energy equation (\ref{fKN}), we parametrically plot $F-T$ curves as presented in Fig. \ref{fig: FT PLOT KN RN KERR}. The dependence on $C$ is eliminated by introducing the relative temperature and relative free energy.

There are three plots shown. Plot (a) and (b) represent $F(T)$ curves for the RN-AdS and Kerr black hole limits, respectively. These plots match with figures \ref{fig: FT RN} and \ref{fig: FT KERR}.

Figure (c) describes the behavior of the free energy as a function of temperature for the KN-AdS black hole at fixed $J$, $\Bar{Q}$, and $C$. Since the electric charge $\Bar{Q}$ and the angular momentum $J$ are related by the dimensionless parameter, we can plot the curves for different values of $J$ or $\Bar{Q}$. Here, we choose the former.

The characteristic properties of these curves are similar to the RN-AdS and Kerr-AdS black holes in RPS formalism. The curves display a ``swallowtail" shape when $J<J_c$. They start at $T=0$, and the entropy increases with $T$.

Above the critical temperature, $T>T_c$, the curve intersects itself at a point known as the self-intersection point. At this point, there are two branches: one with higher entropy and another with lower entropy. These two branches are connected by an unstable intermediate branch with negative heat capacity. This indicates a first-order phase transition at the self-intersection point between two thermodynamically stable branches.

As we increase the value of $J$, the swallowtail gradually becomes smaller, and the temperature at which the self-intersection approaches the critical temperature $T_c$.

At $J=J_c$, there is a kink in the curve at $T=T_c$, and the first-order phase transition becomes second-order.

For $J>J_c$, the curves become smooth and monotonic, and no phase transition is observed.

\subsubsection{\textit{T-S and F-T process at J=0} }
The above discussion applies to non-zero electric charge and angular momentum. Now, we analyze the same scenario for $J=0$. We consider two cases: one for $\Bar{Q}=0$, which corresponds to the Schwarzschild-AdS black hole, and the other for $\Bar{Q}\neq0$, which corresponds to the RN-AdS case.

The results should align with the analysis presented in sections \ref{Schwarzschild AdS} and \ref{Fixed phi TS process}.

Setting the value of $J=0$ into equations (\ref{TKN1}) and (\ref{FKN1}), we obtain:
\begin{equation}
    T=\frac{-\pi^4\Bar{Q}^4+\pi^2(C^2+2\Bar{Q}^2)S^2+4C\pi S^3+3S^4}{4l\sqrt{C}\pi^{3/2}S^{3/2}(\pi^2\Bar{Q}^2+C\pi S+S^2)},
\end{equation}
\begin{equation}
    F=\frac{3\pi^2\Bar{Q}^2+C\pi S-S^2}{4l\pi^{3/2}\sqrt{CS}}.
\end{equation}
Now at $\Bar{Q}=0$, 
\begin{equation}
    T=\frac{C\pi+3S}{4l\pi^{3/2},\sqrt{CS}},\,\,\,\,\,\,\,\,F=\frac{C\pi S-S^2}{4l\pi^{3/2}\sqrt{CS}}.
\end{equation}
At $S=S_{min}$, we find the minimum values of $T$ and $F$, which are given by
\begin{equation}
    T_{min}=\frac{\sqrt{3}}{2l\pi},\,\,\,\,\,\,\,F_{min}=\frac{C}{6\sqrt{3}l},\,\,\,\,\,\,\,\,S_{min}=\frac{\pi C}{3}.
\end{equation}
The results align precisely with equations (\ref{FTKERR}). The corresponding $T-S$ and $F-T$ behaviors are depicted in Fig. \ref{fig: FT MIN KERR}.

Now, we aim to analyze the case of non-zero $\Bar{Q}$. To study the iso-voltage process, we need to express $T$ in terms of $\Bar{\Phi}, C$, and $S$. To achieve this, we utilize equation (\ref{PHIKN1}) to express $\Bar{Q}$ in terms of $\Bar{\Phi}$, resulting in,
\begin{equation}
    T(\Bar{\Phi},S,C)=\frac{C(\pi-l^2\pi\Phi^2)+3S}{4l\pi^{3/2}\sqrt{CS}}.
\end{equation}
From this expression, we find that $T$ has a minimum at $S_{min}$. The values of $S_{min}$ are given by
\begin{equation}
\begin{split}\label{TMINKN}
    & T_{min}=\frac{\sqrt{3-3l^2\Bar{\Phi}^2}}{2l\pi},\\
     & S_{min}=\frac{C\pi}{3}(1-l^2\Bar{\Phi}^2).
\end{split}
\end{equation}
So, here we revert to the results of the RN-AdS case (\ref{SMINTMINRN}). The corresponding behavior of $T-S$ curves is presented in Fig. \ref{fig: TSFT RN}. 

\subsubsection{\texorpdfstring{\textit{T-S process at fixed} $\Omega, \Bar{\Phi}$}{T-S process at fixed \Omega, \Bar{\Phi}}} 

\begin{figure}
    \centering
    \begin{subfigure}[b]{0.48\textwidth}
    \centering
    \includegraphics[width=\textwidth]{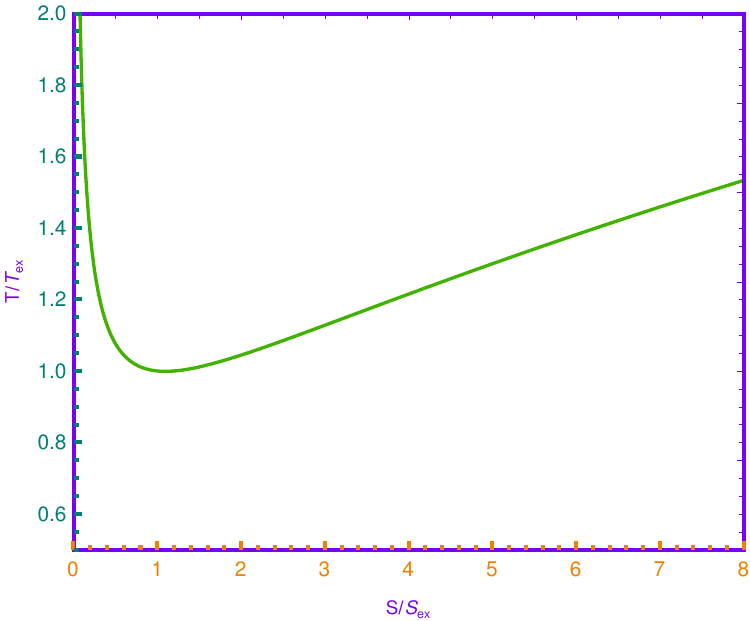}
         \caption{}
         \label{fig: TTEX1 VS SSEX1}
     \end{subfigure}
     \hfill
     \begin{subfigure}[b]{0.48\textwidth}
    \centering
    \includegraphics[width=\textwidth]{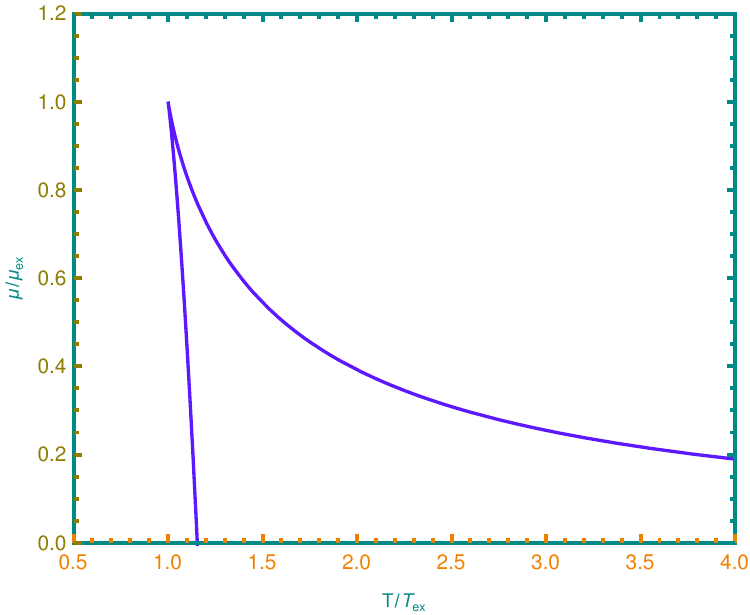}
    \caption{}
     \label{fig: MUMUEX VS SSEX}
     \end{subfigure}
     
    \caption{Figures (a) and (b) display $T-S$ and $\mu-T$ curves, respectively, at fixed $\Omega$ and $\Bar{\Phi}$, utilizing equations (\ref{tbar1}) and (\ref{mubar1}). We conducted a numerical analysis to determine the critical point. The curves are plotted for $\Omega=10$, $\Bar{\Phi}=1$, and $l=0.001$.}
     \label{fig: TMUEX CURVES}
\end{figure}

In the preceding sections, we delved into the thermodynamics of KN-AdS black holes in the canonical ensemble. Now, we shift our focus to the thermodynamic phase structure in the grand canonical ensemble, where $\Omega$ and $\Phi$ are fixed. Initially, we need to express $T$ in terms of $\Omega$ and $\Phi$. Utilizing equations (\ref{OMEGAKN1}) and (\ref{PHIKN1}), we can express $J$ and $\Bar{Q}$ in terms of $\Omega$ and $\Bar{\Phi}$ as follows
\begin{equation}\label{JKNFINAL}
     J=\frac{\Omega l S^{3/2}\sqrt{C\pi+S}(S+C(\pi+l^2\pi\Bar{\Phi}^2)-l^2S\Omega^2)}{2\sqrt{C}\pi^{3/2}(C\pi+S-l^2 S\Omega^2 )},
\end{equation}
and
\begin{equation}\label{QKNFINAL}
    \Bar{Q}=\frac{l\Bar{\Phi}\sqrt{CS(C\pi+S)}}{\sqrt{\pi(C\pi+S-l^2 S \Omega^2)}}.
\end{equation}
Next, substitute the obtained expressions for $J$ and $\Bar{Q}$ into equation (\ref{TKN1}) and utilize the scaling $S=\mathcal{S}C$. This yields the temperature in the following form,
\begin{equation}\label{tbar1}
    \begin{split}
&T(\mathcal{S},\Bar{\Phi},\Omega)\\&=\frac{(\pi+\mathcal{S})^2(-3\mathcal{S}-\pi+\pi l^2\Bar{\Phi}^2)+l^2\mathcal{S}(3\pi^2+6\mathcal{S}^2+\pi\mathcal{S}(9-l^2\Bar{\Phi}^2))\Omega^2-l^4\mathcal{S}^2(2\pi+3\mathcal{S})\Omega^4}{4l\sqrt{\pi^3 \mathcal{S}(\pi+\mathcal{S})(\pi+\mathcal{S}-l^2\mathcal{S}\Omega^2)^3}}.
    \end{split}
\end{equation}
Thanks to the scaling property, the above expression becomes independent of $C$.
\begin{figure}
    \centering
    \begin{subfigure}[b]{0.48\textwidth}
    \centering
    \includegraphics[width=\textwidth]{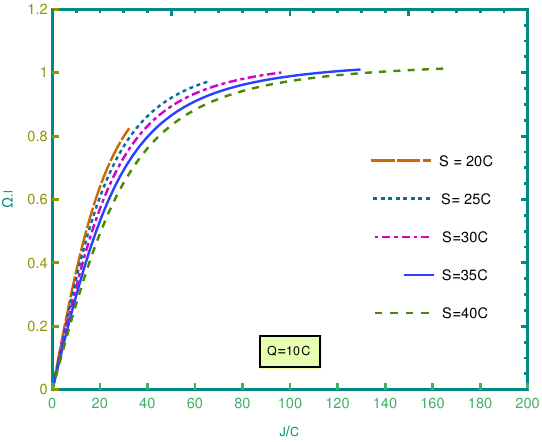}
         \caption{}
         \label{fig: OMEGA J FIXED Q}
     \end{subfigure}
     \hfill
     \begin{subfigure}[b]{0.48\textwidth}
    \centering
    \includegraphics[width=\textwidth]{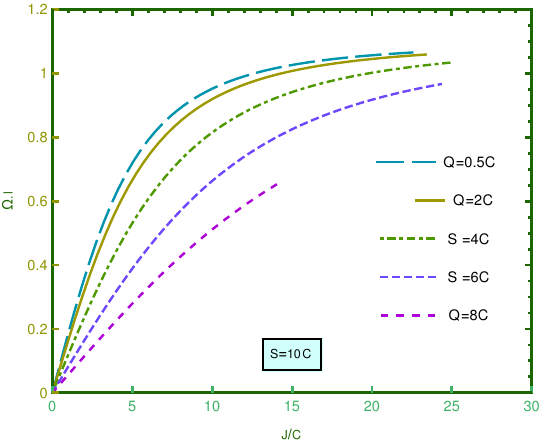}
    \caption{}
     \label{fig: OMEGA J FIXED S}
     \end{subfigure}

     \caption{$\Omega-J$ curves at fixed (a) $\Bar{Q}$ and (b) $S$.}
     \label{fig: OMEGA J KN}
\end{figure}

We deduce from the relationship,
\begin{equation}
    \frac{\partial T}{\partial\mathcal{S}}\Big|_{(\Omega,\Bar{\Phi})}=0.
\end{equation}
 We infer from this relation that there exists a single extremum (minimum) for $T$, located at ($T_{ex}, S_{ex}$). The analytical expressions for $T_{ex}$ and $S_{ex}$ are complex and not particularly informative, so we resort to numerical methods. Nonetheless, in appropriate limits, we can reproduce the corresponding results for RN (\ref{SMINTMINRN}) and Kerr (\ref{SMINKERR5}). Introducing relative parameters as $\Bar{t}={T}/{T_{ex}}$ and $\Bar{s}={\mathcal{S}}/{\mathcal{S}{ex}}$, we plot the curves of ${T}/{T{ex}}$ versus ${S}/{S_{ex}}$ for fixed $\Omega$ and $\Phi$ in figure \ref{fig: TTEX1 VS SSEX1}. The minimum temperature of the process is denoted as $T_{min}$. Above this lowest temperature, there exist two black hole states with equal $\Omega$ and $\Phi$ but different sizes, corresponding to different entropies. The larger black holes, or higher entropic states, are thermodynamically favored over the smaller ones. Consequently, under small perturbations, a smaller black hole transitions to a larger black hole with the same angular velocity and electric potential. This phenomenon can be described as a non-equilibrium phase transition.

 We are now intrigued by exploring the $\mu-T$ behavior at fixed $\Omega$ and $\Bar{\Phi}$, as it provides valuable insights into the $T-S$ process within the grand canonical ensemble. The chemical potential $\mu$ is particularly significant, representing the Gibbs free energy per unit central charge. By substituting equations (\ref{JKNFINAL}) and (\ref{QKNFINAL}) into equation (\ref{muKN1}), we derive the expression for $\mu(S,\Bar{\Phi},\Omega)$ as
\begin{equation}\label{mubar1}
\begin{split}
      &\mu(S,\Bar{\Phi},\Omega)\\&=-\frac{\mathcal{S}((\pi+\mathcal{S})^2(\mathcal{S}+\pi(-1+l^2\Bar{\Phi}^2))-l^2\mathcal{S}(-\pi^2+\pi \mathcal{S}+2\mathcal{S}^2+l^2\pi(2\pi+\mathcal{S})\Bar{\Phi}^2)\Omega^2+l^4\mathcal{S}^3\Omega^4)}{4l\sqrt{\pi^3 \mathcal{S}(\pi+\mathcal{S})(\pi+\mathcal{S}-l^2\mathcal{S}\Omega^2)^3}}.
\end{split}
\end{equation}
After numerical analysis, we observe a peak in the $\mu-T$ curve positioned at ($\mu_{ex}$, $S_{ex}$).

We employ equations (\ref{tbar1}) and (\ref{mubar1}) to parametrically plot the $\Bar{m}-\Bar{t}$ curve in Fig. \ref{fig: MUMUEX VS SSEX}, where $\Bar{m}=\frac{\mu}{\mu_{ex}}$.

\subsubsection{\texorpdfstring{$\Omega-J$ \textit{process at fixed} $S$}{\Omega-J process at fixed S}} 

By employing dimensional analysis, we observe that the extensive variables are proportional to $C$. Hence, we can express them as $S=\mathcal{S}C$, $J=\mathcal{J}C$, and $\Bar{Q}=\mathcal{\Bar{Q}}C$. Substituting these redefined variables into equation (\ref{OMEGAKN1}), we find
\begin{equation}\label{OMEGAPLOT}
    \Omega=\frac{2\pi^{3/2}\mathcal{J}(\pi+\mathcal{S})}{l\sqrt{\mathcal{S}(4\pi^3\mathcal{J}^2(\pi+\mathcal{S})+(\pi^2\mathcal{\Bar{Q}}^2+\pi\mathcal{S}+\mathcal{S}^2)^2)}}.
\end{equation}
It's worth noting that by expressing the intensive quantities in terms of $C$, the expression of $\Omega$ becomes independent of $C$. 

The limit of $J$ given in equation (\ref{LIMITOFJKN}) can be expressed in terms of $\mathcal{J}$ as follows
\begin{equation}
    \mathcal{J}_{max}=\frac{\sqrt{3\mathcal{S}^4+4\pi\mathcal{S}^3+2\pi^2\mathcal{S}^2\mathcal{\Bar{Q}}^2+\pi^2\mathcal{S}^2-\pi^4\mathcal{\Bar{Q}}^4}}{2\pi^2}.
\end{equation}

Using equation (\ref{OMEGAPLOT}), we can plot the $\Omega-J$ curves, as depicted in figure \ref{fig: OMEGA J KN}. When $\Bar{Q}=0$, we recover the Kerr-AdS case, and the corresponding curves are presented in Fig. \ref{fig: OMEGA J curve}. In Fig. \ref{fig: OMEGA J KN}, two scenarios are depicted: curves presented in Fig. \ref{fig: OMEGA J FIXED Q} and Fig. \ref{fig: OMEGA J FIXED S} show the $\Omega-J$ process for fixed $\Bar{Q}$ and different $S$, and for fixed $S$ but different $\Bar{Q}$, respectively. Both plots display a monotonic behavior, starting at the origin and ending at $J=J_{max}$. No phase transition on the $\Omega-J$ plane is observed.

\subsubsection{\texorpdfstring{$\mu-C$\textit{ process at fixed} $S, \Omega, \Bar{\Phi}$}{\mu-C process at fixed S, \Omega, \Bar{\Phi}}} 

\begin{figure}
    \centering
     \centering
    \includegraphics[width=0.50\textwidth]{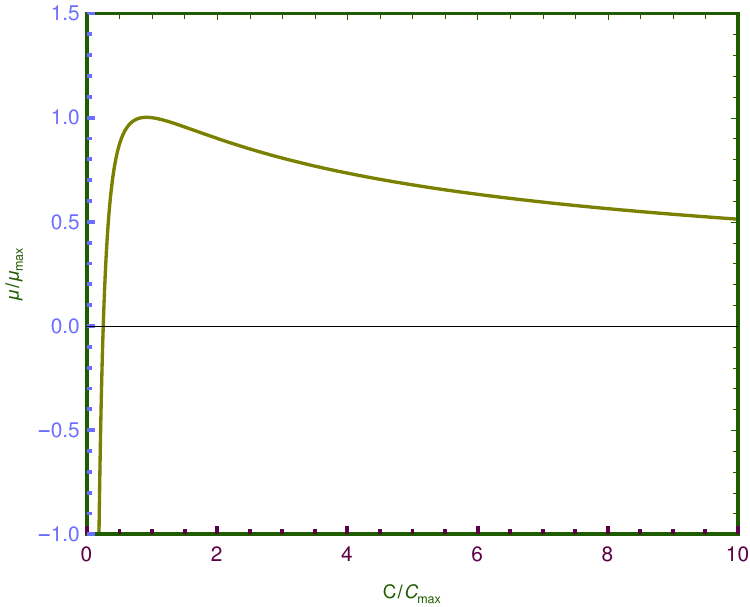}
      \caption{$\mu-C$ curve at $\Omega=10, S=10, \Bar{\Phi}=1\,\, \text{and}\,\, l=0.001$ is depicted here.}
     \label{fig: MU C KN PLOT}
\end{figure}

Here, we delve into another intriguing thermodynamic process: the Gibbs free energy per unit central charge $\mu$ versus $C$. We express $\mu$ as a function of $\Omega$ and $\Bar{\Phi}$ using equations (\ref{muKN1}), (\ref{JKNFINAL}), and (\ref{QKNFINAL}) as follows
\begin{equation}
    \begin{split}
        \mu(\Omega, \Bar{\Phi}, S)
        =-\frac{\sqrt{S}((C\pi+S)^2(S+C\pi(-1+l^2\Bar{\Phi}^2))-\mathcal{A})}{4l\sqrt{C^3\pi^3(C\pi+S)(C\pi+S-l^2S\Omega^2)^3}},
    \end{split}
\end{equation}
 where, $\mathcal{A}=l^2S(-C^2\pi^2+C\pi S+2S^2+l^2C\pi(2C\pi+S)\Bar{\Phi}^2)\Omega^2+l^4S^3\Omega^4$.

From the above equation, it's evident that there is a maximum of $\mu$ at fixed $S$, $\Omega$, and $\Bar{\Phi}$. The location of this maximum is denoted as $(\mu_{max}, C_{max})$. We define dimensionless variables as $c = C/C_{max}$ and $\Bar{m} = \mu/\mu_{max}$. By fixing values for $S$, $\Omega$, and $\Bar{\Phi}$, we can plot the $\mu-C$ curves, as shown in Fig. \ref{fig: MU C KN PLOT}. These curves exhibit similar behavior to those observed in RN-AdS and Kerr-AdS black holes.

\section{Conclusion}
In this paper, we commenced by revisiting restricted phase space thermodynamics, examining two scenarios: RN-AdS and Kerr black holes. Subsequently, we delved into the critical phenomena associated with various thermodynamic processes for KN-AdS black holes. Notably, in this formalism, the mass of the black hole serves as the internal energy, with the AdS radius remaining fixed, distinguishing it from the extended phase space thermodynamics. Our analysis revealed the adherence of the first law, Euler relation, and Gibbs-Duhem equation for KN-AdS black holes within this formalism.

The KN-AdS black hole, characterized by non-zero electric charge and angular momentum, introduces two distinct parameters. By introducing the dimensionless parameter $\beta$, we numerically tackled the first and second-order partial differential equations to identify critical points. Remarkably, we derived a highly accurate fitting formula with minimal parameters, demonstrating excellent agreement when compared with the RN-AdS and Kerr AdS cases. Our investigation encompassed the examination of $T-S$, $\Omega-J$, and $\mu-C$ processes in both the canonical and grand canonical ensembles, uncovering several intriguing phase transitions. We observed that below a critical value of angular momentum, both the $T-S$ and $F-T$ curves exhibit nonmonotonic behavior and a swallowtail shape, respectively. This characteristic behavior is indicative of a van der Waals-like first-order phase transition, which transitions to second-order at the critical temperature. In addition, we delved into the grand canonical ensemble, focusing on the $T-S$ process at fixed $\Omega$ and $\Bar{\Phi}$. Our precise numerical analysis revealed the existence of two distinct black hole states above a minimal temperature, each corresponding to the same angular momentum or electric potential. The states with higher entropy were found to be more preferred, leading to a jump under a small disturbance. We also examined the variation of $\Omega$ with $J$, observing no phase transitions in these curves. Lastly, we explored the $\mu-C$ curves, identifying a maximal chemical potential below which a first-order phase transition occurs. 

By leveraging the equivalence between thermal states in dual conformal field theory and the thermodynamics of AdS black holes, we can derive various thermodynamic models. These models include extended phase space, central charge, and restricted phase space, each corresponding to different variables within the theory. In extended-phase space thermodynamics, the cosmological constant is treated as a dynamical variable, identified with thermodynamic pressure. Alternatively, the first law of thermodynamics can be formulated by considering both the cosmological constant and central charge as variables. In the restricted phase space formalism, the cosmological constant is fixed, but the gravitational constant is treated as a variable. Despite these different approaches, they all exhibit the same type of phase transition, namely the Van der Waals-like phase transition, as explored in various studies. Recent research has examined this thermodynamic behavior from a topological perspective \cite{Zhang_2023, Wei_2022, Wei1_2022}, suggesting that the topological equivalence of these models may underlie the consistent occurrence of similar phase transitions across different models.

Extending this investigation to higher dimensions and exploring higher curvature gravity theories could yield valuable insights. By broadening the scope of our analysis, we may uncover new phenomena and further elucidate the thermodynamic behavior of black holes in diverse gravitational theories.

\section*{Acknowledgements}
The author expresses gratitude to Amit Ghosh, Avirup Ghosh, and Pritam Nanda for their invaluable suggestions and insightful comments. Additionally, the author acknowledges the assistance of MATHEMATICA in obtaining several of the results presented herein.

\bibliographystyle{JHEP}
\bibliography{reference}

\providecommand{\href}[2]{#2}\begingroup\raggedright\begin{thebibliography}{10}

\bibitem{Bardeen:1973gs}
J.~M. Bardeen, B.~Carter and S.~W. Hawking, \emph{{The Four laws of black hole mechanics}}, \href{http://dx.doi.org/10.1007/BF01645742}{\emph{Commun. Math. Phys.} {\bf 31} (1973) 161--170}.

\bibitem{PhysRevD.7.2333}
J.~D. Bekenstein, \emph{Black holes and entropy}, \href{http://dx.doi.org/10.1103/PhysRevD.7.2333}{\emph{Phys. Rev. D} {\bf 7} (Apr, 1973) 2333--2346}.

\bibitem{Hawking:1975vcx}
S.~W. Hawking, \emph{{Particle Creation by Black Holes}}, \href{http://dx.doi.org/10.1007/BF02345020}{\emph{Commun. Math. Phys.} {\bf 43} (1975) 199--220}.

\bibitem{Hawking:1982dh}
S.~W. Hawking and D.~N. Page, \emph{{Thermodynamics of Black Holes in anti-De Sitter Space}}, \href{http://dx.doi.org/10.1007/BF01208266}{\emph{Commun. Math. Phys.} {\bf 87} (1983) 577}.

\bibitem{witten1998antide}
E.~Witten, \emph{Anti-de sitter space, thermal phase transition, and confinement in gauge theories},  1998.

\bibitem{Birmingham_1999}
D.~Birmingham, \emph{Topological black holes in anti-de sitter space}, \href{http://dx.doi.org/10.1088/0264-9381/16/4/009}{\emph{Classical and Quantum Gravity} {\bf 16} (Jan., 1999) 1197–1205}.

\bibitem{Emparan_1999}
R.~Emparan, \emph{Ads/cft duals of topological black holes and the entropy of zero-energy states}, \href{http://dx.doi.org/10.1088/1126-6708/1999/06/036}{\emph{Journal of High Energy Physics} {\bf 1999} (June, 1999) 036–036}.

\bibitem{Chamblin_1999}
A.~Chamblin, R.~Emparan, C.~V. Johnson and R.~C. Myers, \emph{Holography, thermodynamics, and fluctuations of charged ads black holes}, \href{http://dx.doi.org/10.1103/physrevd.60.104026}{\emph{Physical Review D} {\bf 60} (Oct., 1999) }.

\bibitem{Louko_1996}
J.~Louko and S.~N. Winters-Hilt, \emph{Hamiltonian thermodynamics of the reissner—nordström—anti-de sitter black hole}, \href{http://dx.doi.org/10.1103/physrevd.54.2647}{\emph{Physical Review D} {\bf 54} (Aug., 1996) 2647–2663}.

\bibitem{Mitra_1998}
P.~Mitra, \emph{Entropy of extremal black holes in asymptotically anti-de sitter spacetime}, \href{http://dx.doi.org/10.1016/s0370-2693(98)01161-7}{\emph{Physics Letters B} {\bf 441} (Nov., 1998) 89–95}.

\bibitem{Pe_a_1999}
C.~S. Peça and J.~P.~S. Lemos, \emph{Thermodynamics of reissner–nordström–anti-de sitter black holes in the grand canonical ensemble}, \href{http://dx.doi.org/10.1103/physrevd.59.124007}{\emph{Physical Review D} {\bf 59} (May, 1999) }.

\bibitem{Caldarelli_1999}
M.~M. Caldarelli, G.~Cognola and D.~Klemm, \emph{Thermodynamics of kerr-newman-ads black holes and conformal field theories}, \href{http://dx.doi.org/10.1088/0264-9381/17/2/310}{\emph{Classical and Quantum Gravity} {\bf 17} (Dec., 1999) 399–420}.

\bibitem{Kastor_2009}
D.~Kastor, S.~Ray and J.~Traschen, \emph{Enthalpy and the mechanics of ads black holes}, \href{http://dx.doi.org/10.1088/0264-9381/26/19/195011}{\emph{Classical and Quantum Gravity} {\bf 26} (Sept., 2009) 195011}.

\bibitem{Dolan_2011}
B.~P. Dolan, \emph{The cosmological constant and black-hole thermodynamic potentials}, \href{http://dx.doi.org/10.1088/0264-9381/28/12/125020}{\emph{Classical and Quantum Gravity} {\bf 28} (May, 2011) 125020}.

\bibitem{Dolan1_2011}
B.~P. Dolan, \emph{Pressure and volume in the first law of black hole thermodynamics}, \href{http://dx.doi.org/10.1088/0264-9381/28/23/235017}{\emph{Classical and Quantum Gravity} {\bf 28} (Nov., 2011) 235017}.

\bibitem{Cai_2013}
R.-G. Cai, L.-M. Cao, L.~Li and R.-Q. Yang, \emph{P-v criticality in the extended phase space of gauss-bonnet black holes in ads space}, \href{http://dx.doi.org/10.1007/jhep09(2013)005}{\emph{Journal of High Energy Physics} {\bf 2013} (Sept., 2013) }.

\bibitem{Kubiz_k_2017}
D.~Kubizňák, R.~B. Mann and M.~Teo, \emph{Black hole chemistry: thermodynamics with lambda}, \href{http://dx.doi.org/10.1088/1361-6382/aa5c69}{\emph{Classical and Quantum Gravity} {\bf 34} (Mar., 2017) 063001}.

\bibitem{KUBZANAK}
D.~Kubizňák and R.~B. Mann, \emph{P-v criticality of charged ads black holes}, {\emph{Journal of High Energy Physics} (2012) }.

\bibitem{Altamirano_2013}
N.~Altamirano, D.~Kubizňák and R.~B. Mann, \emph{Reentrant phase transitions in rotating anti–de sitter black holes}, \href{http://dx.doi.org/10.1103/physrevd.88.101502}{\emph{Physical Review D} {\bf 88} (Nov., 2013) }.

\bibitem{Johnson_2014}
C.~V. Johnson, \emph{Holographic heat engines}, \href{http://dx.doi.org/10.1088/0264-9381/31/20/205002}{\emph{Classical and Quantum Gravity} {\bf 31} (Oct., 2014) 205002}.

\bibitem{Altamirano_2014}
N.~Altamirano, D.~Kubizňák, R.~B. Mann and Z.~Sherkatghanad, \emph{Kerr-ads analogue of triple point and solid/liquid/gas phase transition}, \href{http://dx.doi.org/10.1088/0264-9381/31/4/042001}{\emph{Classical and Quantum Gravity} {\bf 31} (Feb., 2014) 042001}.

\bibitem{Wei_2014}
S.-W. Wei and Y.-X. Liu, \emph{Triple points and phase diagrams in the extended phase space of charged gauss-bonnet black holes in ads space}, \href{http://dx.doi.org/10.1103/physrevd.90.044057}{\emph{Physical Review D} {\bf 90} (Aug., 2014) }.

\bibitem{Frassino_2023}
A.~M. Frassino, J.~F. Pedraza, A.~Svesko and M.~R. Visser, \emph{Higher-dimensional origin of extended black hole thermodynamics}, \href{http://dx.doi.org/10.1103/physrevlett.130.161501}{\emph{Physical Review Letters} {\bf 130} (Apr., 2023) }.

\bibitem{Visser_2022}
M.~R. Visser, \emph{Holographic thermodynamics requires a chemical potential for color}, \href{http://dx.doi.org/10.1103/physrevd.105.106014}{\emph{Physical Review D} {\bf 105} (May, 2022) }.

\bibitem{Kastor_2014}
D.~Kastor, S.~Ray and J.~Traschen, \emph{Chemical potential in the first law for holographic entanglement entropy}, \href{http://dx.doi.org/10.1007/jhep11(2014)120}{\emph{Journal of High Energy Physics} {\bf 2014} (Nov., 2014) }.

\bibitem{Karch_2015}
A.~Karch and B.~Robinson, \emph{Holographic black hole chemistry}, \href{http://dx.doi.org/10.1007/jhep12(2015)073}{\emph{Journal of High Energy Physics} {\bf 2015} (Dec., 2015) 1–15}.

\bibitem{Maity_2017}
R.~Maity, P.~Roy and T.~Sarkar, \emph{Black hole phase transitions and the chemical potential}, \href{http://dx.doi.org/10.1016/j.physletb.2016.12.004}{\emph{Physics Letters B} {\bf 765} (Feb., 2017) 386–394}.

\bibitem{ahmed2023holographic}
M.~B. Ahmed, W.~Cong, D.~Kubizňák, R.~B. Mann and M.~R. Visser, \emph{Holographic dual of extended black hole thermodynamics},  2023.

\bibitem{Gao_2022}
Z.~Gao and L.~Zhao, \emph{Restricted phase space thermodynamics for ads black holes via holography}, \href{http://dx.doi.org/10.1088/1361-6382/ac566c}{\emph{Classical and Quantum Gravity} {\bf 39} (Mar., 2022) 075019}.

\bibitem{Gong_2023}
T.-F. Gong, J.~Jiang and M.~Zhang, \emph{Holographic thermodynamics of rotating black holes}, \href{http://dx.doi.org/10.1007/jhep06(2023)105}{\emph{Journal of High Energy Physics} {\bf 2023} (June, 2023) }.

\bibitem{Cong_2022}
W.~Cong, D.~Kubizňák, R.~B. Mann and M.~R. Visser, \emph{Holographic cft phase transitions and criticality for charged ads black holes}, \href{http://dx.doi.org/10.1007/jhep08(2022)174}{\emph{Journal of High Energy Physics} {\bf 2022} (Aug., 2022) }.

\bibitem{Gaok_2022}
Z.~Gao, X.~Kong and L.~Zhao, \emph{Thermodynamics of kerr-ads black holes in the restricted phase space}, \href{http://dx.doi.org/10.1140/epjc/s10052-022-10080-y}{\emph{The European Physical Journal C} {\bf 82} (Feb., 2022) }.

\bibitem{PhysRevLett.130.181401}
M.~B. Ahmed, W.~Cong, D.~Kubiz\ifmmode~\check{n}\else \v{n}\fi{}\'ak, R.~B. Mann and M.~R. Visser, \emph{Holographic dual of extended black hole thermodynamics}, \href{http://dx.doi.org/10.1103/PhysRevLett.130.181401}{\emph{Phys. Rev. Lett.} {\bf 130} (May, 2023) 181401}.

\bibitem{Cvetic_1999}
M.~Cvetic and S.~S. Gubser, \emph{Phases of r-charged black holes, spinning branes and strongly coupled gauge theories}, \href{http://dx.doi.org/10.1088/1126-6708/1999/04/024}{\emph{Journal of High Energy Physics} {\bf 1999} (Apr., 1999) 024–024}.

\bibitem{PhysRevD.93.084015}
S.-W. Wei, P.~Cheng and Y.-X. Liu, \emph{Analytical and exact critical phenomena of $d$-dimensional singly spinning kerr-ads black holes}, \href{http://dx.doi.org/10.1103/PhysRevD.93.084015}{\emph{Phys. Rev. D} {\bf 93} (Apr, 2016) 084015}.

\bibitem{Cheng_2016}
P.~Cheng, S.-W. Wei and Y.-X. Liu, \emph{Critical phenomena in the extended phase space of kerr-newman-ads black holes}, \href{http://dx.doi.org/10.1103/physrevd.94.024025}{\emph{Physical Review D} {\bf 94} (July, 2016) }.

\bibitem{PhysRevD.108.044045}
M.~S. Ali, S.~G. Ghosh and A.~Wang, \emph{Thermodynamics of kerr-sen-ads black holes in the restricted phase space}, \href{http://dx.doi.org/10.1103/PhysRevD.108.044045}{\emph{Phys. Rev. D} {\bf 108} (Aug, 2023) 044045}.

\bibitem{Zhang_2023}
M.~Zhang and J.~Jiang, \emph{Bulk-boundary thermodynamic equivalence: a topology viewpoint}, \href{http://dx.doi.org/10.1007/jhep06(2023)115}{\emph{Journal of High Energy Physics} {\bf 2023} (June, 2023) }.

\bibitem{Wei_2022}
S.-W. Wei and Y.-X. Liu, \emph{Topology of black hole thermodynamics}, \href{http://dx.doi.org/10.1103/physrevd.105.104003}{\emph{Physical Review D} {\bf 105} (May, 2022) }.

\bibitem{Wei1_2022}
S.-W. Wei, Y.-X. Liu and R.~B. Mann, \emph{Black hole solutions as topological thermodynamic defects}, \href{http://dx.doi.org/10.1103/physrevlett.129.191101}{\emph{Physical Review Letters} {\bf 129} (Oct., 2022) }.

\end{thebibliography}\endgroup

\end{document}